\RequirePackage{fix-cm}
\documentclass[twocolumn,epjc3]{svjour3} 
\smartqed  
\RequirePackage{amssymb}
\RequirePackage{amsmath}
\RequirePackage{array} 
\RequirePackage{epsfig}
\RequirePackage{graphics}
\RequirePackage{colortbl}
\RequirePackage{hhline}
\RequirePackage{multirow}
\begin{document}


\title{Central exclusive diffractive production of two-pions from continuum and decays of resonances in the Regge-eikonal model 
}


\author{R.A.~Ryutin \thanksref{e1,addr1}
}

\thankstext{e1}{e-mail: Roman.Rioutine@cern.ch}


\institute{{\small NRC ``Kurchatov Institute'' - Institute for High Energy Physics, Protvino {\it 142 281}, Russia} \label{addr1}
}

\date{
}

\maketitle

\begin{abstract}
Calculations of central exclusive diffractive production (CEDP) of two pions via continuum and resonance mechanism are presented in the Regge-eiko\-nal
approach. Data from STAR, ISR, CDF and CMS were analysed
and compared with theoretical description. Preliminary extraction of $f_0(500)$, $f_0(980)$ and $f_2(1270)$ couplings to Pomeron, and also $\rho$ meson contribution from photoproduction mechanism to the CEDP cross-section are 
considered. We show possible nuances and problems of calculations and pros\-pects of investigations at present and future hadron colliders.
\end{abstract} 
%
%


\section*{Introduction}

In previous papers~\cite{myCEDP1},\cite{myCEDP2} general
properties and calculations of the Central Exclusive Diffractive
Production (CEDP) were considered. It was shown, especially
in~\cite{myCEDP2}, that diffractive patterns (differential cross-sections) of CEDP play significant role in model verification.

In~\cite{myCEDPpipiContinuum} we considered the low mass CEDP 
(LM CEDP) with production of two pion continuum. Here we expand
this analysis and add contributions from CEDP of $f_0(500)$, $f_0(980)$, $f_2(1270)$ and exclusive vector meson photoproduction (EVMP) of $\rho_0$ meson to this process, taking into account also the interference 
between continuum and resonance mechanisms.

CEDP of two pions is one of 
the basic ``standard candles'' for LM CEDP. Why do we need exact calculations and predictions for this process?
\begin{itemize}
\item Di-pion LM CEDP is the tool for the investigation
of hadronic resonances (like $f_2$ or $f_0$), since one of the basic hadronic decay modes for these resonances is the two pion one. We can extract different
couplings of these resonances to reggeons (Pomeron, Odderon etc) to understand their nature (structure and the interaction mechanisms). 
\item We can use LM CEDP to fix the procedure of calculations of ``rescattering'' (unitarity) corrections. In the case of di-pion LM CEDP there
are several kinds of corrections, in the proton-proton, pion-proton and also pion-pion 
channels.
 \item The pion is the most fundamental particle in the strong interactions, and LM CEDP gives us a powerful tool to go deep inside its properties, especially to investigate the form factor and scattering amplitudes for the off-shell pion.
 \item LM CEDP has rather large cross-sections. It is very important for an exclusive process, since in the special low luminocity runs (of the LHC) we need more time to get enough statistics.
 \item As was proposed in~\cite{myCEDPpipiContinuum},\cite{mySD}, it 
 is possible to extract some reggeon-hadron cross-sections. In the case of single and
  double dissociation it was the Pomeron-proton one. Here, in the LM CEDP of the di-
  pion we can ana\-ly\-ze properties of the Pomeron-Pomeron to pion-pion exclusive 
  cross-section.
 \item Diffractive patterns of this process are very sensitive to different
approaches (subamplitudes, form factors, unitarization, reggeization procedure), especially differential cross-sections in $t$ and $\phi_{pp}$ (azi\-mu\-thal angle between final protons), and also $M_{\pi\pi}$ dependence. That is why this process is used to verify different models of diffraction.
\item All the above items are additional advantages provided by the LM CEDP of two pions, which has usual properties of CEDP: clear  signature  with  two final protons and two  large  rapidity  gaps  (LRG)~\cite{LRG1},\cite{LRG2}  and the  possibility  to  use  the  “missing  mass  method”~\cite{MMM}.
\end{itemize}

Processes of the LM CEDP were calculated in some other works~\cite{CEDPw1}-\cite{CEDPw6} which are devoted to most popular models
for the LM CEDP of di-mesons, where authors have considered phenomenological, nonperturbative, perturbative and mixed approaches in Reggeon-Reggeon collision subprocess.  Nuances of some approa\-ches were analysed in the introduction
 of~\cite{myCEDPpipiContinuum}.

\begin{figure}[hbt!]
\begin{center}
\includegraphics[width=0.49\textwidth]{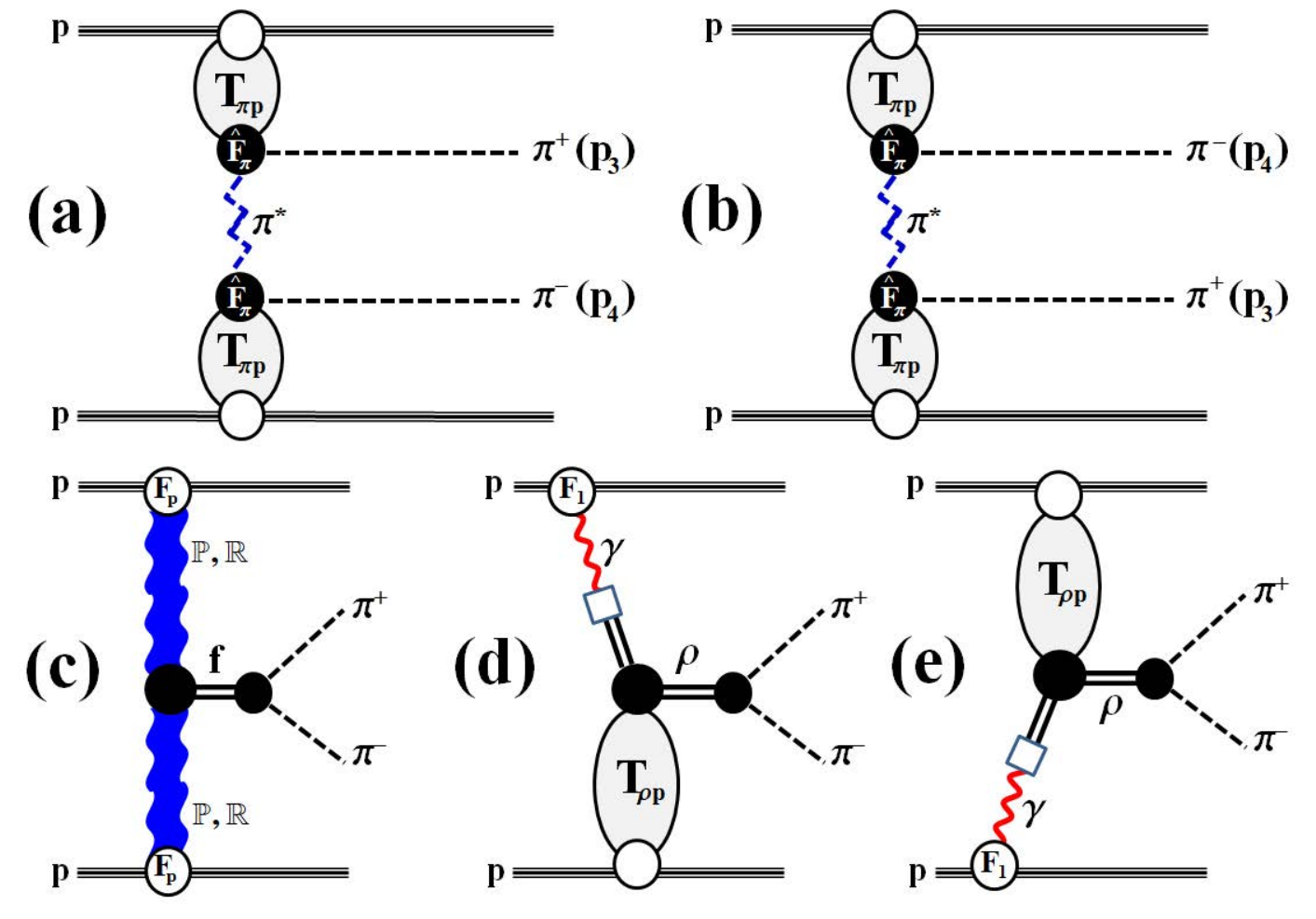}
\caption{\label{fig3:MY4CASES} Primary amplitudes (absorptive corrections are not shown) of the process of double pion LM CEDP $p+p\to p+\pi^++\pi^-+p$ in the Regge-eikonal approach for continuum (a),(b), LM CEDP of $f_0$ and $f_2$ resonances (c) and EVMP of $\rho_0$ meson (d),(e) with subsequent decay to $\pi^+\pi^-$. (a),(b): central part of the diagram is the primary continuum CEDP amplitude, where $T_{\pi^{\pm}p}$ are full elastic pion-proton amplitudes, and the reggeized off-shell pion propagator depicted as dahsed zigzag line. This is the most applicable case from four approaches considered in~\cite{myCEDPpipiContinuum}. (c): central part of the diagram contains Pomeron-Pomeron-resonance fusion with subsequent decay to pions, propagator is taken in the Breit-Wigner approximation. (d),(e):  central part of the diagram contains full EVMP amplitude with $\rho_0$ production and decay. Off-shell pion form factor on (a),(b) and other suppression form-factors (in the Pomeron-Pomeron-f or the pion-pion-($f_0$,$f_2$,$\rho_0$ vertices) ) on (c),(d),(e) are presented as a black circles.}
\end{center}
\end{figure}

In this article we consider the case, depicted
in Fig.~\ref{fig3:MY4CASES}, and show how it can describe
the data from ISR~\cite{ISRdata1}, \cite{ISRdata2},
STAR~\cite{STARdata1}-\cite{STARdata5}, CDF~\cite{CDFdata1},\cite{CDFdata2},
CMS~\cite{CMSdata1}-\cite{CMSdata4} collaborations.

In the first part of the present work we introduce the framework
for calculations of double pion LM CEDP (kinematics, amplitudes, differential
cross-sections) in the Regge-eikonal approach, wich was considered in details in~\cite{myCEDPpipiContinuum}. Here we take reggeized full (RF) case from four approaches of~\cite{myCEDPpipiContinuum} which is the best in the data description.

Also we discuss some nuances
of the calculations, which we should take into account in 
further more accurate analysis (elastic amplitudes for virtual 
particles, off-shell pion form factor, pion-pion elastic
amplitude at low energies, nonlinearity of the 
pion trajectory, tensorial couplings and spin effects).

In the second part we analyse the experimental data on the process
at different energies, extract some couplings of resonances to Pomeron, find the best approach and make some predictions for LHC experiments.

To avoid complicated expressions in the main text, all the basic formulae are placed
to Appendixes.
  
\begin{figure}[hbt!]
\begin{center}
\includegraphics[width=0.49\textwidth]{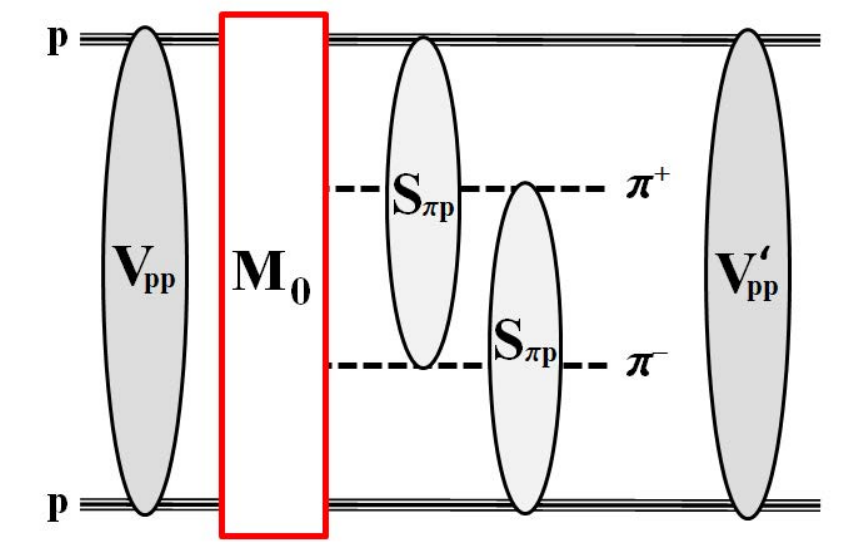}
\caption{\label{fig:UM0} Full unitarized amplitude of the process of double pion LM CEDP $p+p\to p+\pi^++\pi^-+p$. Proton-proton rescatterings in the initial and final states are depicted as $V_{pp}$ and $V'_{pp}$-blobes correspondingly, and pion-proton rescattering corrections are also shown as $S_{\pi p}$-blobes. Sum of primary amplitudes $M_0$ from Fig.~\ref{fig3:MY4CASES} are shown as a red rectangle.}
\end{center}
\end{figure}   
  
\section{General framework for calculations of LM CEDP}

LM CEDP is the first exclusive two to four process which is driven basically by the Pomeron-Pomeron fusion subprocess. It serves as a clear process for 
investigations of resonances like $f_0(500)$, $f_0(980)$, $f_2(1270)$ and others with masses less than $3$~GeV. At the noment, for low central di-pion masses it is a huge problem to use perturbative approach, that is why we apply the Regge-eikonal method for all the calculations. For proton-proton and proton-pion elastic amplitudes we use the model of~\cite{godizovpp}, \cite{godizovpip}, which describe all the available experimental data on elastic scattering. For EVMP we use
the similar model~\cite{godizovmesons}.

\subsection{Components of the framework}

LM CEDP process can be calculated in the following scheme (see Fig.~\ref{fig3:MY4CASES}):
\begin{enumerate}

\item
We calculate the primary amplitudes of the processes, which are depicted as central parts of diagrams in Fig.~\ref{fig3:MY4CASES}. Here we
consider the case, where the bare off-shell pion propagator in the amplitude for continuum
di-pion production is replaced by the reggeized one
\begin{eqnarray}
{\cal P}_{\pi}(\hat{s},\hat{t})&=&
\left(
\mathrm{ctg}\frac{\pi\alpha_{\pi}(\hat{t})}{2}-\mathrm{i}
\right)\cdot\nonumber\\
&\cdot&\frac{\pi\alpha_{\pi}^{\prime}}{2\phantom{\mbox{\large{I}}^2}\!\!\!\!\mathrm{\Gamma}(1+\alpha_{\pi}(\hat{t}))}
\left(
\frac{\hat{s}}{s_0}
\right)^{\alpha_{\pi}(\hat{t})},
\label{eq:PionPropagatorR}
\end{eqnarray}
where $\hat{s}$ is the di-pion mass squared and $\hat{t}$ is the square of the momentum transfer between a  Pomeron and a pion in the Pomeron-Pomeron fusion process (see Appendix~A for details). 

In the present calculations we use linear pion trajectory $\alpha_{\pi}(\hat{t})=0.7(\hat{t}-m_{\pi}^2)$. Nonlinear case was also
verified, and the difference in the final result is not significant.

We use also full eikonalized expressions for proton-proton, pion-proton and photon-proton amplitudes, which can be found in the Appendix~B.

\item
After the calculation of the primary LM CEDP amplitudes we have to take into account all possible corrections in proton-proton and proton-pion elastic channels due to the unitarization procedure (so called ``soft survival probability'' or ``rescattering corrections''), which are depicted as $V_{pp}$, $V'_{pp}$ and $S_{\pi p}$ blobes in Fig.~\ref{fig3:MY4CASES}. For proton-proton and proton-pion elastic amplitudes we use the model of~\cite{godizovpp}, \cite{godizovpip} (see Appendix~B). Possible final pion-pion interaction is not 
shown in Fig.~\ref{fig3:MY4CASES}, since we neglect it in the present calculations.

\end{enumerate}

In this article we do not consider so called ``enhanced'' corrections~\cite{CEDPw1}-\cite{CEDPw3}, since they give nonleading contributions in our model due to smallness of the triple Pomeron vertex. Also we have no possible absorptive  corrections in the pion-pion final elastic channel, since the central mass is low, and also there is a lack of data on this process to define parameters of the model. Nevertheless one can consider these corrections, as it was done by some authors recently~\cite{pipiespagnol}, since they could play significant role for masses less than 1~GeV.

Exact kinematics of the two to four process is outlined in Appendix~A.

Here we use the model, presented in Appendix~B for example. You can use another one, which is proved 
to describe well all the available data on proton-proton, proton-pion elastic processes and exclusive $\rho_0$ photoproduction. But it is difficult to find 
now more than a couple of models 
which have more or less predictable 
power (see~\cite{godizovmodelsfault} for detailed discussion). That is why we use the model, which is proved to be good 
in data fitting, especially in the kinematical region of our in\-te\-rest.

\subsection{Continuum di-pion production}

Final expression for the amplitude for the continuum di-pion production 
with proton-pro\-ton and pion-proton ``rescattering'' corrections (see Fig.~\ref{fig3:MY4CASES}(a),(b)) can be written as
\begin{eqnarray}
&& M^U\left( \{ p \}\right) = 
\nonumber\\
&&  
=\int\int 
\frac{d^2\vec{q}}{(2\pi)^2}\frac{d^2\vec{q}^{\prime}}{(2\pi)^2}
\frac{d^2\vec{q}_1}{(2\pi)^2}\frac{d^2\vec{q}_2}{(2\pi)^2}
V_{pp}(s,q^2)
V_{pp}(s^{\prime},q^{\prime 2})\nonumber\\
&& \times\; \left[
S_{\pi^-p}(\tilde{s}_{14},q_1^2)
M_0\left( \{ \tilde{p}\} \right)
S_{\pi^+p}(\tilde{s}_{23},q_2^2)+
(3\leftrightarrow 4)
\right] \label{eq:MU}\\
&& M_0\left( \{ p \}\right)=\phantom{I^{2^{2^2}}}\nonumber\\
&&
=T^{el}_{\pi^+p}(s_{13},t_1)
{\cal P}_{\pi}(\hat{s},\hat{t})
\left[ 
\hat{F}_{\pi}\left( \hat{t}\right)
\right]^2
T^{el}_{\pi^-p}(s_{24},t_2),
\label{eq:M0}
\end{eqnarray}
where functions are defined in~(\ref{eq:Vpp1})-(\ref{eq:Tpipexact}) of Appendix~B, and sets of vectors are
\begin{eqnarray}
&&\{ p \}\equiv \{ p_a,p_b,p_1,p_2,p_3,p_4\}\label{eq:setp}\\
&&\{ \tilde{p}\}\equiv \{ p_a-q,p_b+q; p_1+q^{\prime}+q_1,\nonumber\\
&& \;\;\;\;\;\;\;\;\;\;\;\; p_2-q^{\prime}+q_2,p_3-q_2,p_4-q_1 \} \label{eq:setptild},
\end{eqnarray}
and
\begin{eqnarray}
\tilde{s}_{14}&=&\left( p_1+p_4+q^{\prime} \right)^2,\;
\tilde{s}_{23}=\left( p_2+p_3-q^{\prime} \right)^2,
\label{eq:invarstild}\\
s_{ij}&=&\left( p_i+p_j \right)^2,\;
t_{1,2}=\left( p_{a,b}-p_{1,2} \right)^2,
\label{eq:invars}\\
\hat{s}&=&\left( p_3+p_4 \right)^2,\;
\hat{t}=\left( p_a-p_1-p_3 \right)^2
\label{eq:invarsh}
\end{eqnarray}

Off-shell pion form factor is equal to unity on mass shell $\hat{t}=m_{\pi}^2$ and 
taken as exponential
\begin{equation}
\hat{F}_{\pi}=\mathrm{e}^{(\hat{t}-m_{\pi}^2)/\Lambda_{\pi}^2},
\label{eq:offshellFpi}
\end{equation}
where $\Lambda_{\pi}\sim 1.2$~GeV is taken from 
the fits to LM CEDP of two pions at low energies (see next section). In this paper we use only exponential form, but
it is possible to use other parametrizations (see~\cite{CEDPw1}-\cite{CEDPw6}). Exponential one shows more appropriate
results in the data fitting.

Other functions are defined in Appendix~B. Then we can use the expression~(\ref{eq5:dcsdall}) to calculate
the differential cross-section of the process.

\subsection{CEDP and EVMP of low mass resonances.}

 For $f_0(500)$, $f_0(980)$, $f_2(1270)$ and $\rho_0$ resonances the general unitarized
amplitude (see Fig.~\ref{fig3:MY4CASES}(c),(d)) is similar to the expression~(\ref{eq:MU}), where amplitude $M_0\left( \{ p \}\right)$ is replaced by the corresponding central primary amplitude for the resonance production and futher decay to $\pi^+\pi^-$. 

 For $f$ mesons amplitudes are constructed from pro\-ton-Pomeron form-factors, Pomeron-Pomeron couplings to mesons\footnote{Here we take the simple scalar one for every meson, even for $f_2(1270)$ (multiplied by the leading tensor term), although, as was mentioned in our work~\cite{myWA102}, this vertex can be rather complicated and can give nontrivial contribution to the dependence on the azimuthal
 angle between final protons. But for our goals in this paper, namely, investigation of the di-pion mass distributions, it is rather good approximation.}, off-shell propagators, off-shell form-factors and decay vertices. 
 
 For the EVMP amplitude of $\rho_0$ meson we take the full unitarized photon-proton amplitude contracted with the photon flux, and also with the off-shell meson 
propagator, off-shell form-factors and its decay vertex to pions. Also we have to sum this
amplitude with the symmetric one with $t_1\leftrightarrow t_2$.
 
All the primary amplitudes are presented in the Appendix~C. 
 
\subsection{Nuances of calculations.}

In the next section one can see that there are some difficulties and puzzles 
in the data fitting, which 
have also been presented in other works~\cite{CEDPw4}-\cite{CEDPw6}. In this 
subsection let us discuss some nuances of calculations, which could change the situation. 

We have to pay special attention to amplitudes, where one or more external particles
are off their mass shell. The example of such an amplitude is the
pion-proton one $T_{\pi^+ p}$ ($T_{\pi^- p}$), which is the part of the CEDP 
amplitude (see~(\ref{eq:MU})). For this amplitude in the present paper we use Regge-eikonal model with the
eikonal function in the classical Regge form. And ``off-shell'' condition for one of the pions
is taken into account by additional phenomenological form factor 
$\hat{F}_{\pi}(\hat{t})$. But there are at least two other possibilities.

The first one was considered in~\cite{PetrovOffshell}. For amplitude 
with one particle off-shell the formula
\begin{equation}
T^*(s,b)=\frac{\delta^*(s,b)}{\delta(s,b)}T(s,b)=\frac{\delta^*(s,b)}{\delta(s,b)}\frac{\mathrm{e}^{2\mathrm{i}\delta(s,b)}-1}{2\mathrm{i}}
\label{eq:TstarVAP}
\end{equation}
was used. In our case 
\begin{eqnarray}
&& \delta(s,b) = \delta_{\pi p}(s,b; m_{\pi}^2,m_{\pi}^2,m_p^2,m_p^2),\nonumber\\ 
&& \delta^*(s,b) = \delta_{\pi p}^*(s,b; \hat{t},m_{\pi}^2,m_p^2,m_p^2)\nonumber\\
&& \left.\delta_{\pi p}=\delta^*_{\pi p}\right|_{\hat{t}\to m_{\pi}^2}.
\label{eq:deltasVAP}
\end{eqnarray}
$\delta_{\pi p}$ is the eikonal function (see~(\ref{eq:elamplitudes})). This is similar to the introduction of the additional form factor, but in a more consistent way, which takes into account the 
unitarity condition.

\begin{figure}[hbt!]
\begin{center}
\includegraphics[width=0.45\textwidth]{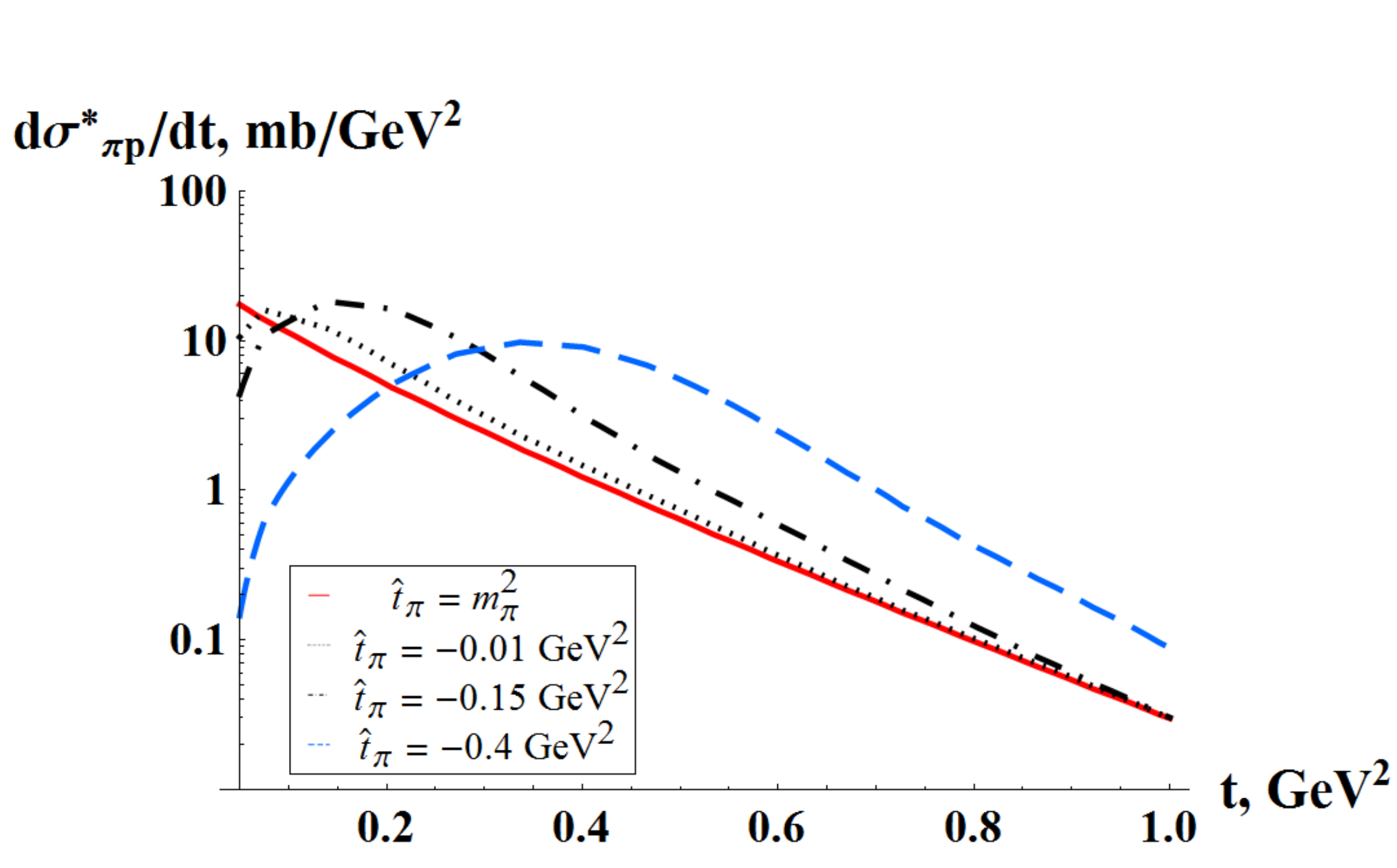}
\caption{\label{fig:offshellpip} Pion-proton on-shell and off-shell elastic differential cross-section (in the
	model of conserved meson currents presented in Appendix~C) for different pion virtualities $\hat{t}_{\pi}$: $m_{\pi}^2$(on-shell),$-0.01$~$GeV^2$,$-0.15$~$GeV^2$,$-0.4$~$GeV^2$ in the covariant approach with conserved currents.}
\end{center}
\end{figure}   

The second one arises from the covariant reggeization
method, which was considered in the Appendix~C of~\cite{myCEDPpipiContinuum}. For the case of conserved hadronic currents
we have definite structure in the Legendre function, which is transformed in a natural way to the case of the off-shell
amplitude. But in this case off-shell amplitude has a specific behaviour at low t values (see Fig.~\ref{fig:offshellpip} and~\cite{myCEDP2} for details). As 
was shown in~\cite{myCEDP2}, unitarity corrections can mask this behavior.

\begin{figure}[hbt!]
	\begin{center}
		\includegraphics[width=0.45\textwidth]{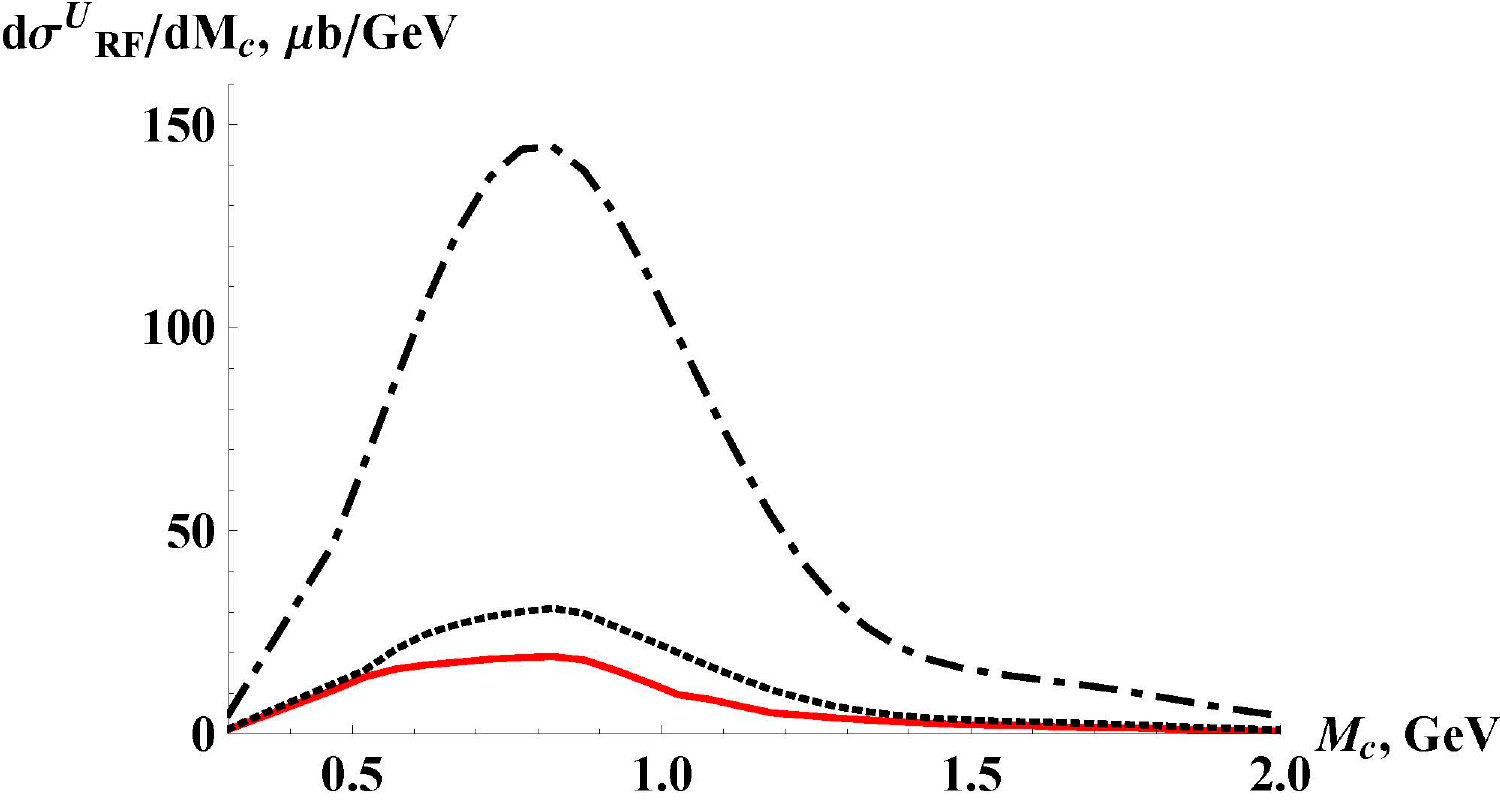}		
		\caption{\label{fig:CMS7RFX} Predictions of the model for the continuum (see Fig.~\ref{fig3:MY4CASES}(a),(b)) at $7$~TeV.  Curves from 
			up to down correspond to the Born term, the amplitude with proton-proton rescattering corrections only and the one with 
			all the corrections (proton-proton and pion-proton). $\Lambda_{\pi}=1.2$~GeV.
		}
	\end{center}
\end{figure}

Since final pion-proton interactions can give rather large
suppression (about 10-20\%, as in Fig.~\ref{fig:CMS7RFX}), in our calculations we use the full amplitude as depicted in  Fig.~\ref{fig:UM0}. 

\section{Data from hadron colliders versus results of calculations}
\label{sec:data}

Our basic task is to extract the fundamental information on the interaction of hadrons from
different cross-sections (``diffractive patterns''):
\begin{itemize}
\item from t-distributions we can obtain size and shape of the interaction region;
\item the distribution on the azimuthal angle between final protons gives quantum numbers of the
produced system (see~\cite{myCEDP2},\cite{myWA102} and references therein);
\item from $M_c$ (here $M_c=M_{\pi\pi}$) dependence and its influence on t-dependence we can make some conclusions about
the interaction at different space-time scales and interrelation between them. Also we can extract couplings of reggeons to different resonances.
\end{itemize}

Process $p+p\to p+\pi+\pi+p$  is the first ``standard candle'', which we can use to estimate other CEDP processes. In this section we consider
the experimental data on the process and make an attempt to extract the information on couplings of different resonances to Pomeron.

\begin{figure}[h!]
\begin{center}
a)\includegraphics[width=0.35\textwidth]{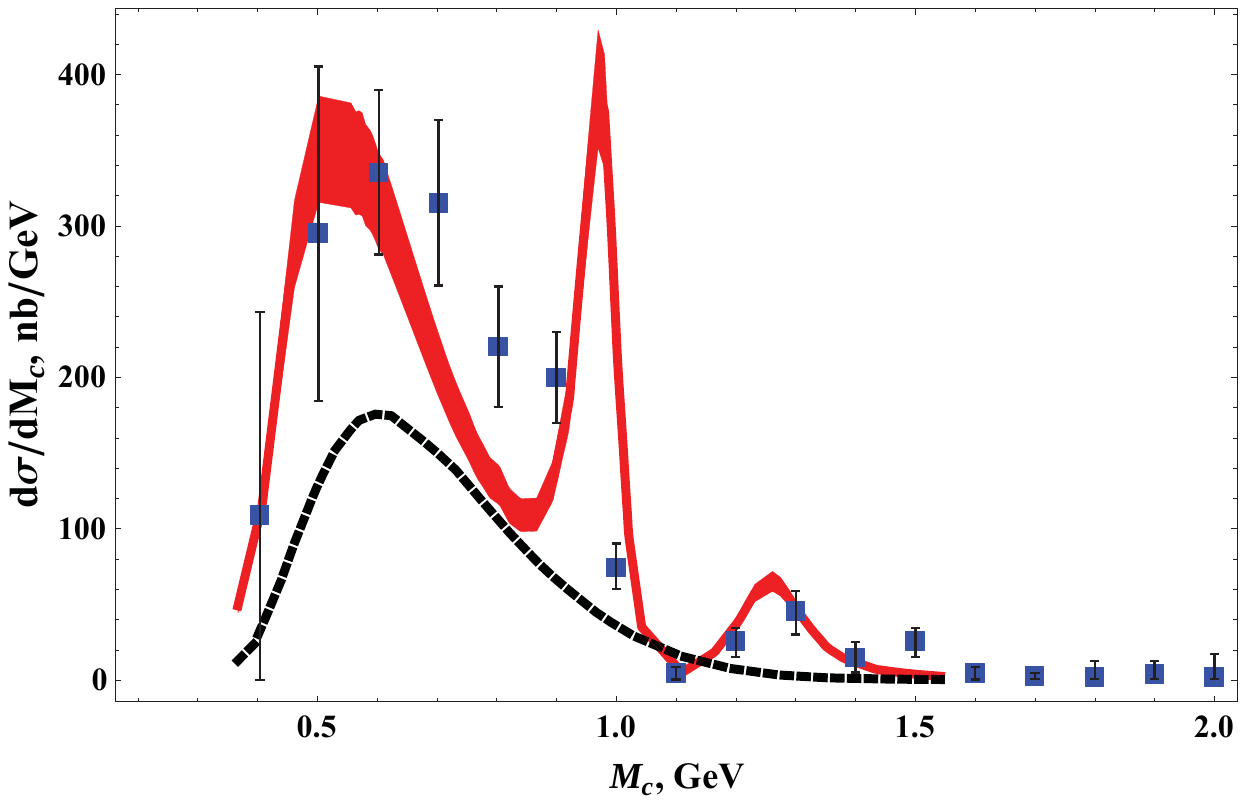}\\
b)\includegraphics[width=0.35\textwidth]{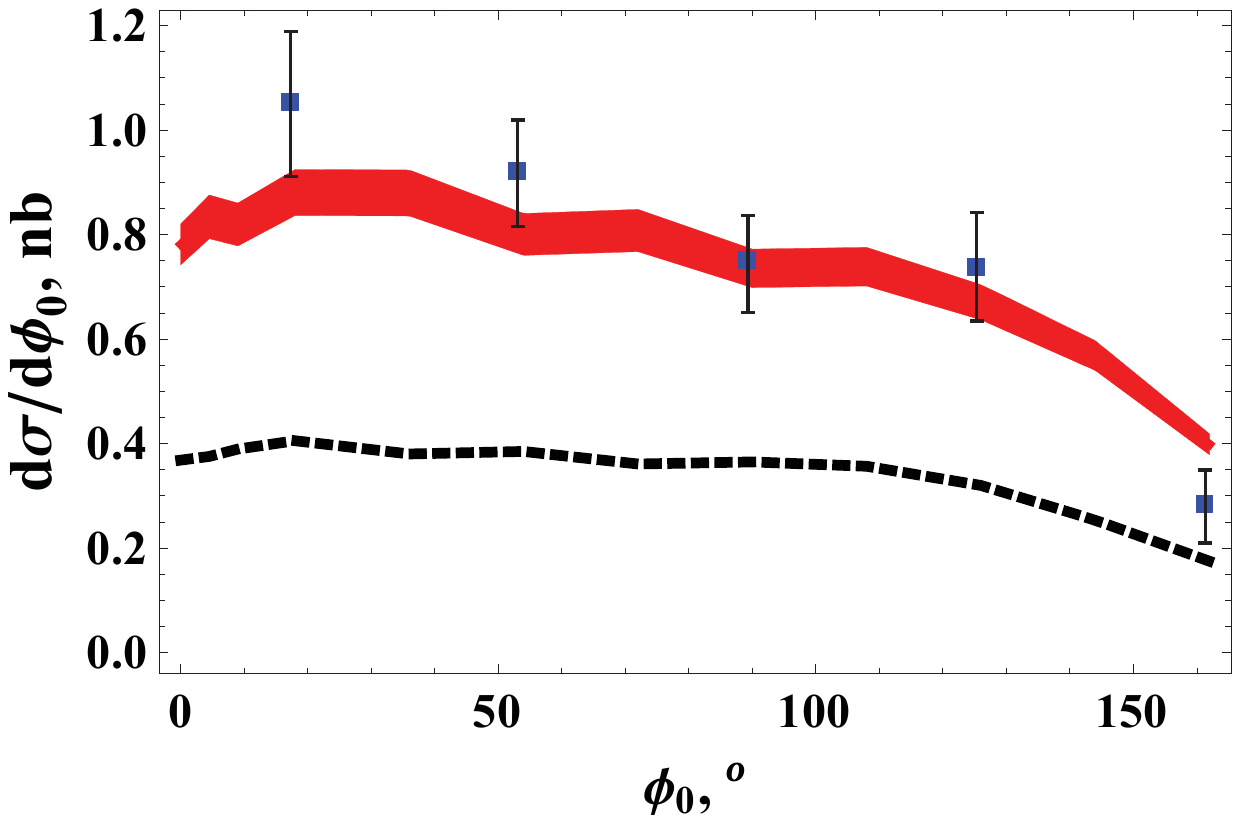}\\
\caption{\label{fig:starOLD} The data on the process $p+p\to p+\pi^++\pi^-+p$ 
at $\sqrt{s}=200$~GeV (STAR collaboration~\cite{STARdata1},\cite{STARdata2}): (a) $|\eta_{\pi}|<1$, $|\eta_{\pi\pi}|<2$, $p_{T\pi}>0.15$~GeV, $0.005<-t_{1,2}<0.03$~GeV$^2$; (b) plus additional cut $M_c < 1$~GeV.  Curves correspond to $\Lambda_{\pi}=1.2$~GeV in the off-shell pion form factor~(\ref{eq:offshellFpi}) and couplings from~(\ref{PPfcouplings}).
Solid upper curves correspond to the sum of all amplitudes and dashed lower ones represent the continuum contribution. Thickness of the solid curves corresponds to the errors of Monte-Carlo calculations.
}
\end{center}
\end{figure}   

\begin{figure}[h!]
\begin{center}
a)\includegraphics[width=0.35\textwidth]{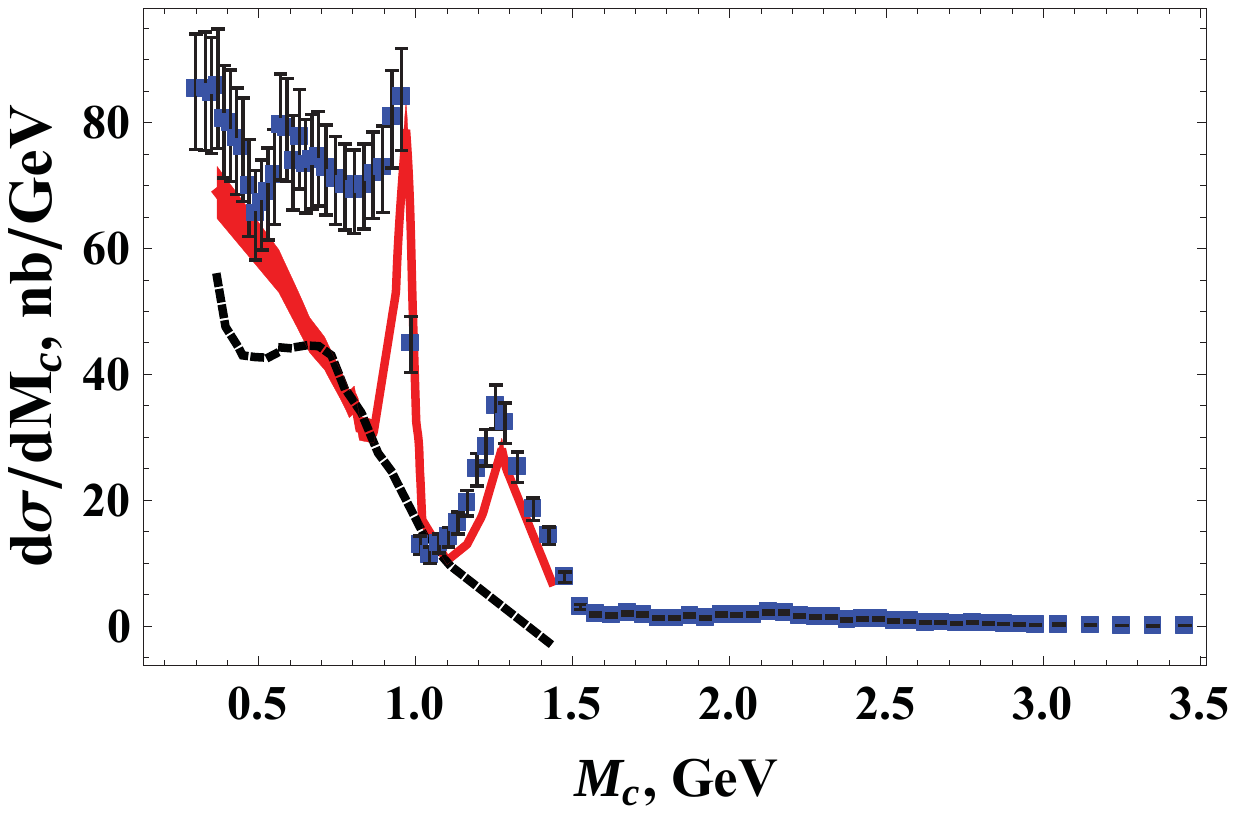}\\
b)\includegraphics[width=0.35\textwidth]{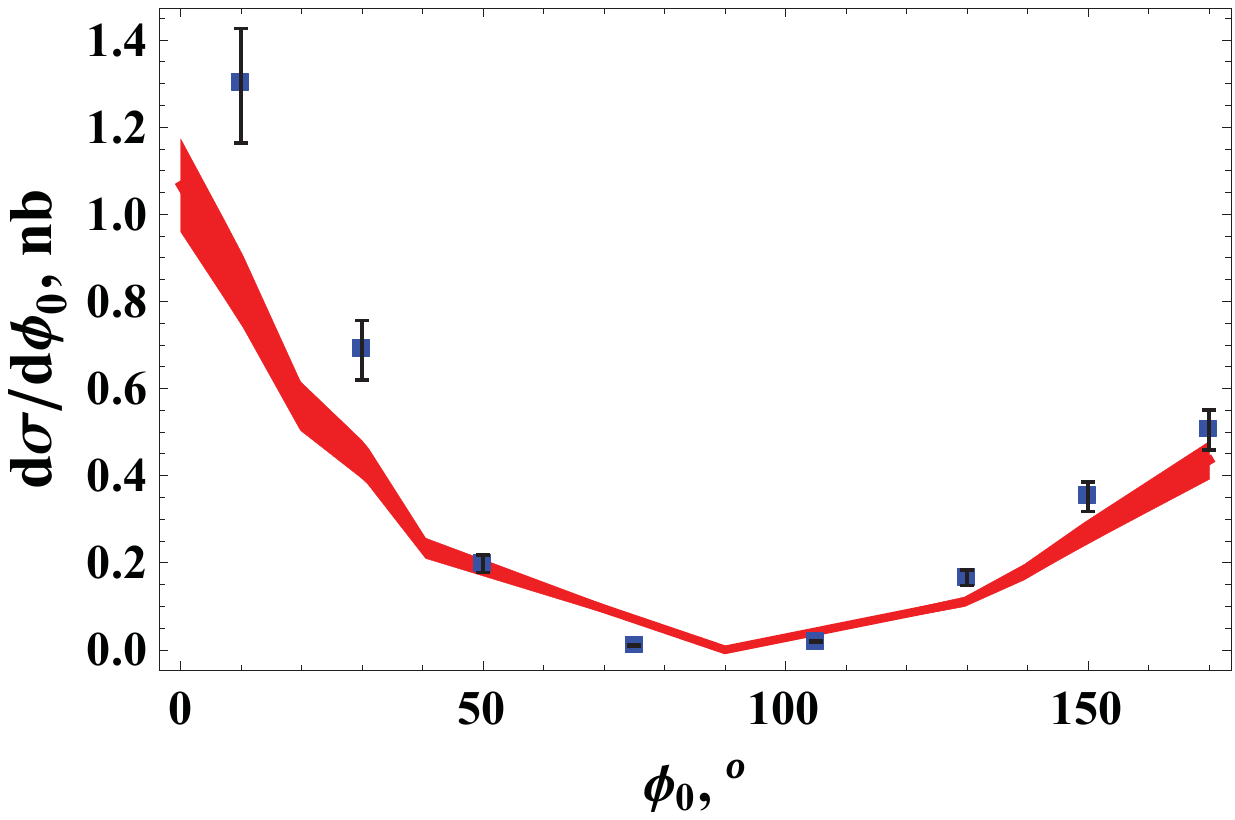}\\
c)\includegraphics[width=0.35\textwidth]{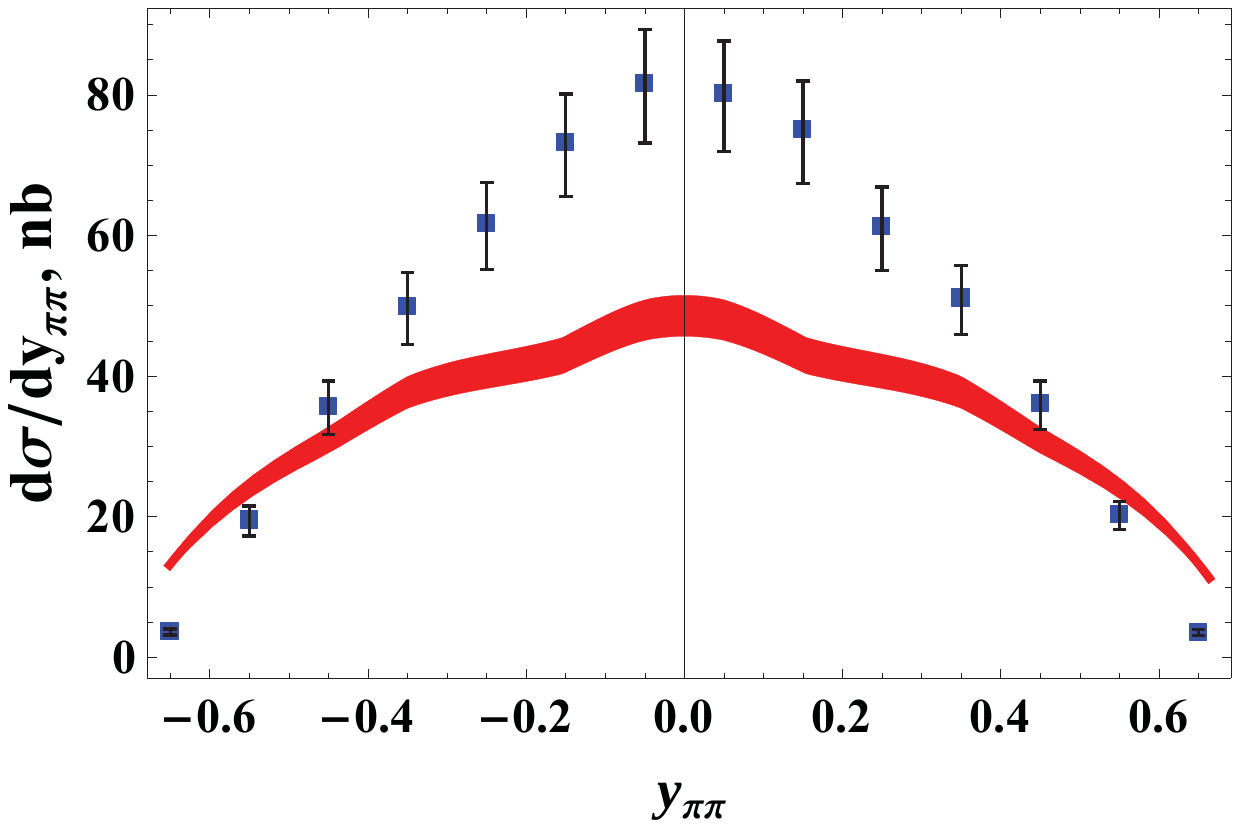}\\
d)\includegraphics[width=0.35\textwidth]{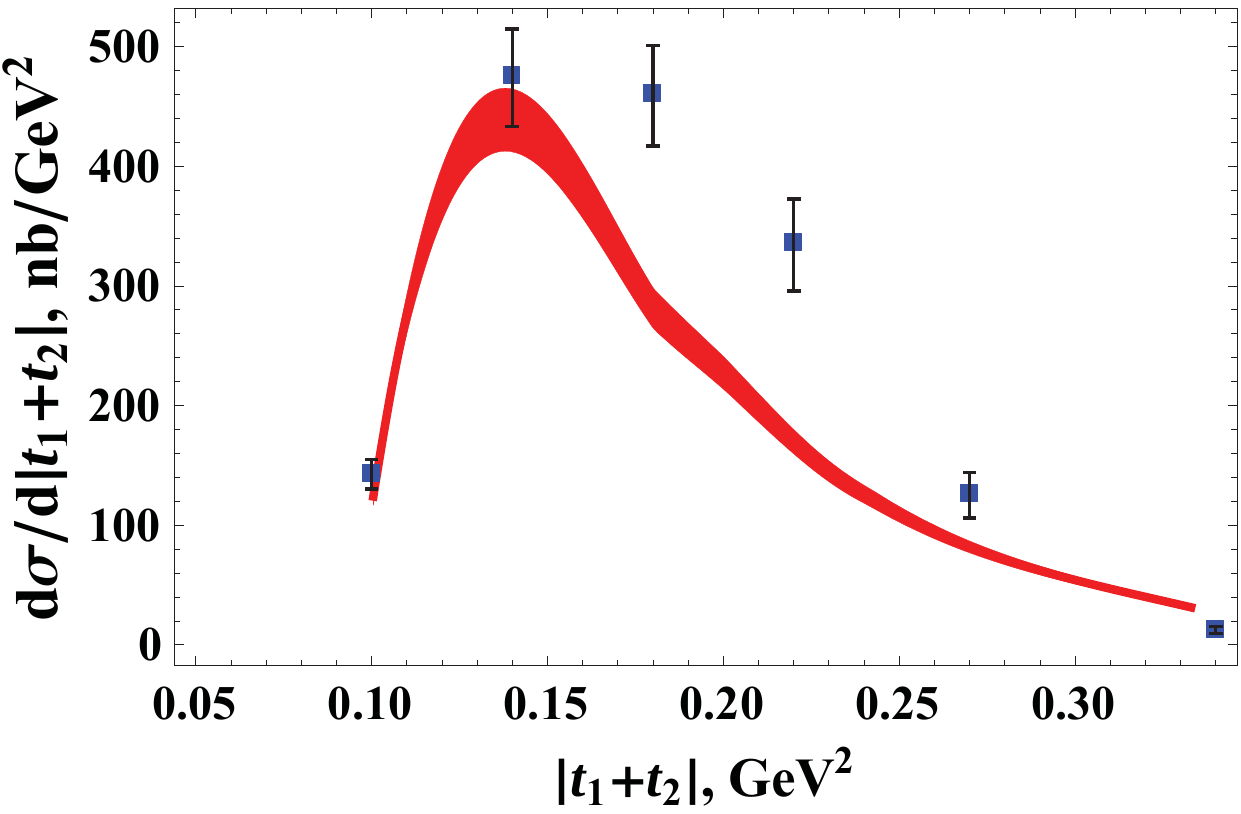}\\
\caption{\label{fig:starNEW} The new data on 
the process $p+p\to p+\pi^++\pi^-+p$ 
at $\sqrt{s}=200$~GeV (STAR collaboration~\cite{STARdata3}-\cite{STARdata4a}): $|\eta_{\pi}|<0.7$, $p_{T\pi}>0.2$~GeV, $p_x>-0.2$~GeV, $0.2$~GeV$<|p_y|<0.4$~GeV, $(p_x+0.3\,\mathrm{GeV})^2+p_y^2<0.25$~GeV$^2$, where $p$ denotes the momenta of final protons.  Curves correspond to $\Lambda_{\pi}=1.2$~GeV in the off-shell pion form factor~(\ref{eq:offshellFpi}) and couplings from~(\ref{PPfcouplings}).
Solid curves correspond to the sum of all amplitudes and the dashed lower one in (a) represents the continuum contribution. Thickness of the solid curves corresponds to the errors of Monte-Carlo calculations.
}
\end{center}
\end{figure}  

\subsection{STAR collaboration data versus model cases}

In this subsection the data of the STAR collaboration~\cite{STARdata1}-\cite{STARdata5} and model curves for continuum and the sum of all cases of Fig.~\ref{fig3:MY4CASES} are presented. In our approach we have several free parameters: $\Lambda_{\pi}$ (for the continuum) and couplings of resonances to Pomeron $g_{\mathbb{PP}f}$, which we can extract from the data. All the distributions are depicted here for $\Lambda_{\pi}=1.2$~GeV and couplings from~(\ref{PPfcouplings}). 

\begin{figure}[h!]
\begin{center}
\includegraphics[width=0.45\textwidth]{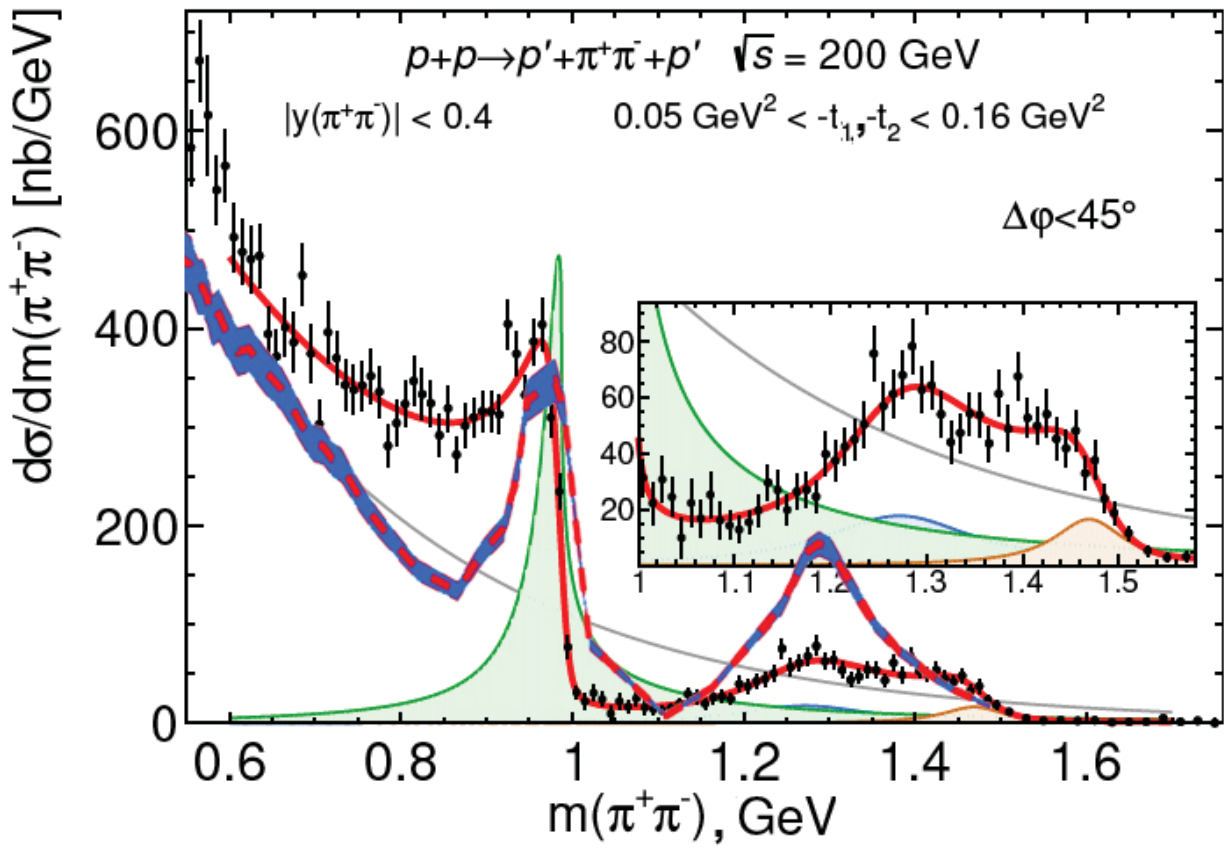}\\
\caption{\label{fig:starNEWFULL} The new data on the process $p+p\to p+\pi^++\pi^-+p$ at $\sqrt{s}=200$~GeV (STAR collaboration~\cite{STARdata3}-\cite{STARdata4a}): $|y_{\pi\pi}|<0.4$, $0.05$~GeV$<|t_{1,2}|<0.16$~GeV$^2$, $\phi_0<45^{o}$.  Dashed curve corresponds to parameters, fixed from the old data from STAR depicted in the Fig.~\ref{fig:starOLD} and  explained in the text: $\Lambda_{\pi}=1.2$~GeV in the off-shell pion form factor~(\ref{eq:offshellFpi}) and couplings from~(\ref{PPfcouplings}).}
\end{center}
\end{figure}   

\begin{figure}[h!]
\begin{center}
\includegraphics[width=0.4\textwidth]{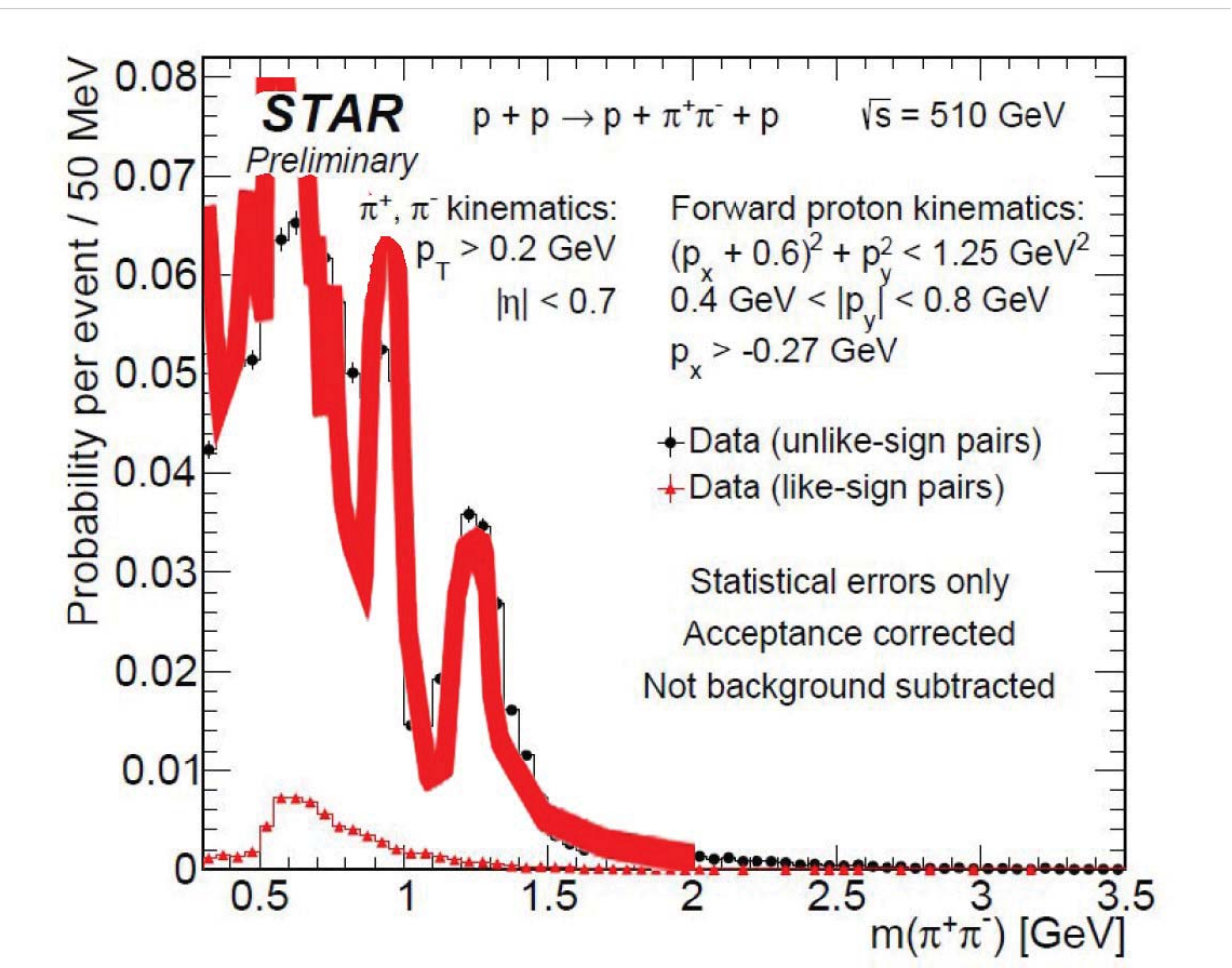}\\
\caption{\label{fig:starNEW510} The new preliminary data on 
the process $p+p\to p+\pi^++\pi^-+p$ 
at $\sqrt{s}=510$~GeV (STAR collaboration~\cite{STARdata5}): $|\eta_{\pi}|<0.7$, $p_{T\pi}>0.2$~GeV, $p_x>-0.27$~GeV, $0.4$~GeV$<|p_y|<0.8$~GeV, $(p_x+0.6\,\mathrm{GeV})^2+p_y^2<1.25$~GeV$^2$, where $p$ denotes the momenta of final protons.  Theoretical solid curve correspond to $\Lambda_{\pi}=1.2$~GeV in the off-shell pion form factor~(\ref{eq:offshellFpi}) and couplings from~(\ref{PPfcouplings}). Fluctuations are due to complex Monte-Carlo integration process, and they can be smoothed by increasing the integration accuracy.
}
\end{center}
\end{figure}   

We begin our analysis by fitting the data from STAR depicted on the Fig.~\ref{fig:starOLD}(a). Then
we make predictions for other available data on this process. 

First of all we have to note a difference between the data and prediction at the same energy. In the Fig.~\ref{fig:starOLD}(b) one can see the azimuthal angle
distribution. Here we could explain some difference, because
our Pomeron-Pomeron-meson couplings are constants, but in reality
they depends on the azimuthal angle. The next case, which is
depicted in Figs.~\ref{fig:starNEW} 
is more interesting. We
see that predictions underestimate the data in the region of $M_c\sim 0.8\pm 0.1$~GeV
as in the Fig.\ref{fig:starOLD}(a). The data are close to predictions in the regions of $f$ mesons. This fact can not be understood at the moment,
since, if we fix parameters using this new STAR data from Figs.~\ref{fig:starNEW},
we will obtain overestimation of the data at
higher energies, which 
looks strange. May be we have to renormalize this part of
the STAR data, or Pomeron-Pomeron-meson couplings have
some complex dependance on $t_{1,2}$ and $\phi_0$. It should be checked
in further investigations. The $\rho$ contribution is small. We also see some differences in Figs.~\ref{fig:starNEW}(b,c,d).

The data for extended kinematical region at 
$\sqrt{s}=200$~GeV is depicted in the Fig.~\ref{fig:starNEWFULL} with
the theoretical curve, which is slightly 
underestimate the data at low masses and overestimate the data at
the peak of $f_2(1270)$. Since these data were obtained by some
extrapolation, it can be considered only qualitatively.

In the Fig.~\ref{fig:starNEW510} we see the similar situation with
overestimation at $M_c\sim 0.7\pm 0.2$~GeV and at the peak of $f_0(980)$.

\subsection{ISR and CDF data versus the model curves}

\begin{figure}[h!]
\begin{center}
a)\includegraphics[width=0.35\textwidth]{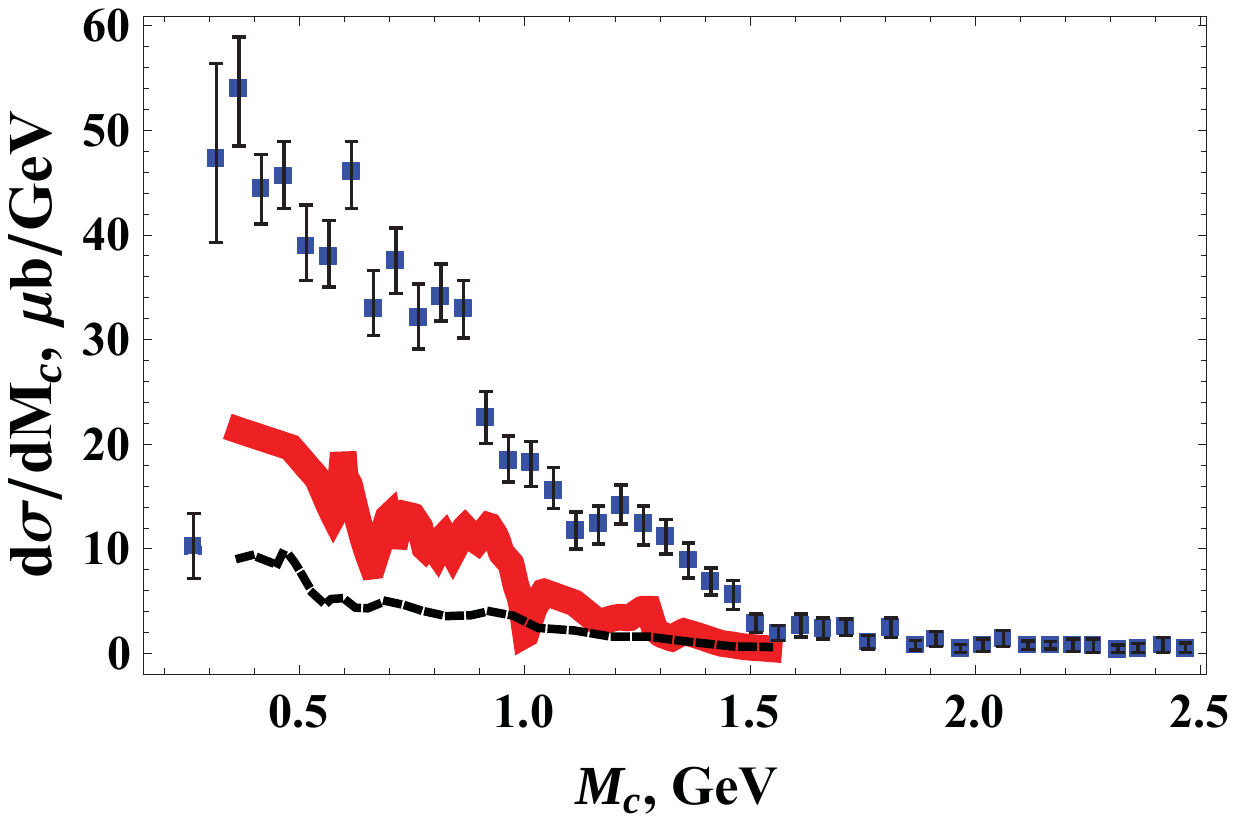}\\
b)\includegraphics[width=0.35\textwidth]{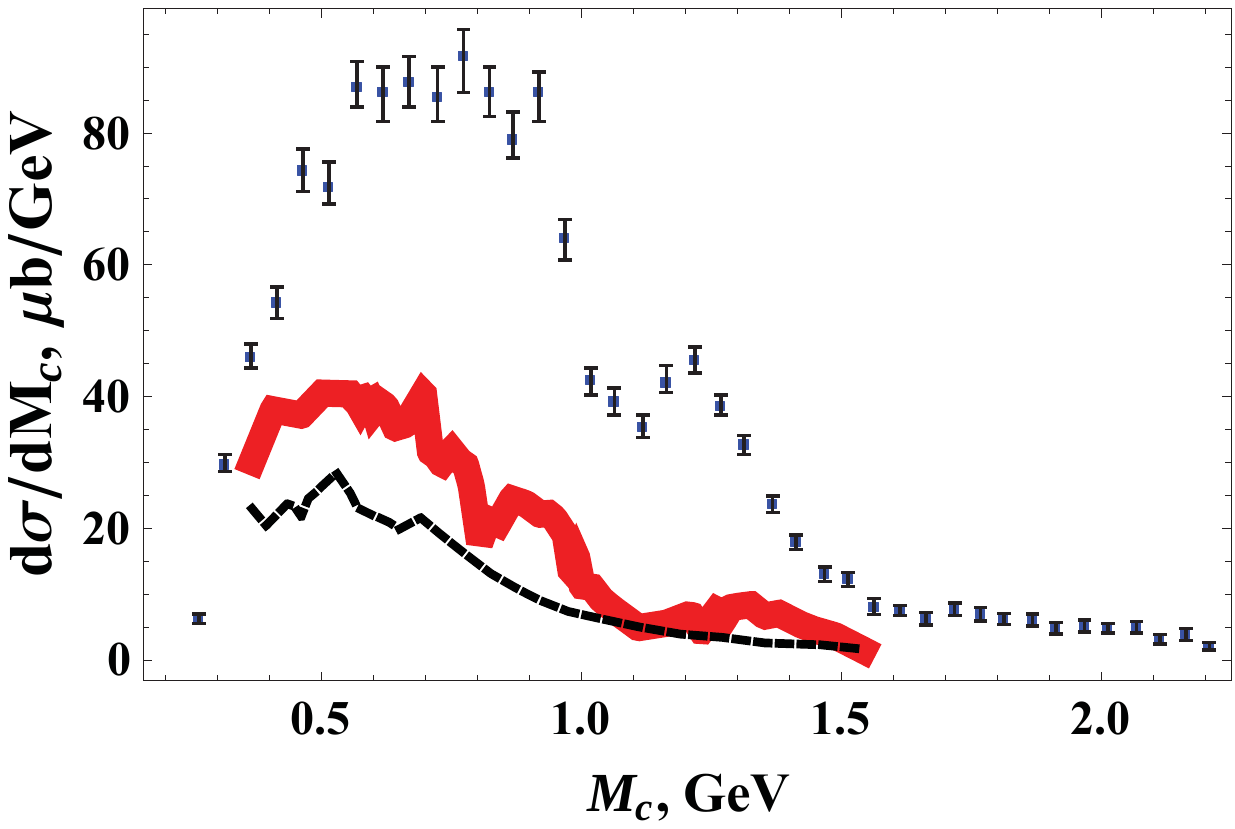}\\
\caption{\label{fig:ISR}  The data on the process $p+p\to p+\pi^++\pi^-+p$  (ISR and ABCDHW collaborations~\cite{ISRdata1},\cite{ISRdata2}): (a)
at $\sqrt{s}=63$~GeV, $|y_{\pi}|<1$, $\xi_p>0.9$; (b) at $\sqrt{s}=62$~GeV, $|y_{\pi}|<1.5$, $\xi_p>0.9$. Curves correspond to $\Lambda_{\pi}=1.2$~GeV in the off-shell pion form factor~(\ref{eq:offshellFpi}) and couplings from~(\ref{PPfcouplings}).
Solid upper curves correspond to the sum of all amplitudes and dashed lower ones represent the continuum contribution. Fluctuations in solid curves are due to complex Monte-Carlo integration process, and they can be smoothed by increasing the integration accuracy. Thickness of the solid curves corresponds to the errors of Monte-Carlo calculations.
}
\end{center}
\end{figure}   

\begin{figure}[hbt!]
\begin{center}
a)\includegraphics[width=0.35\textwidth]{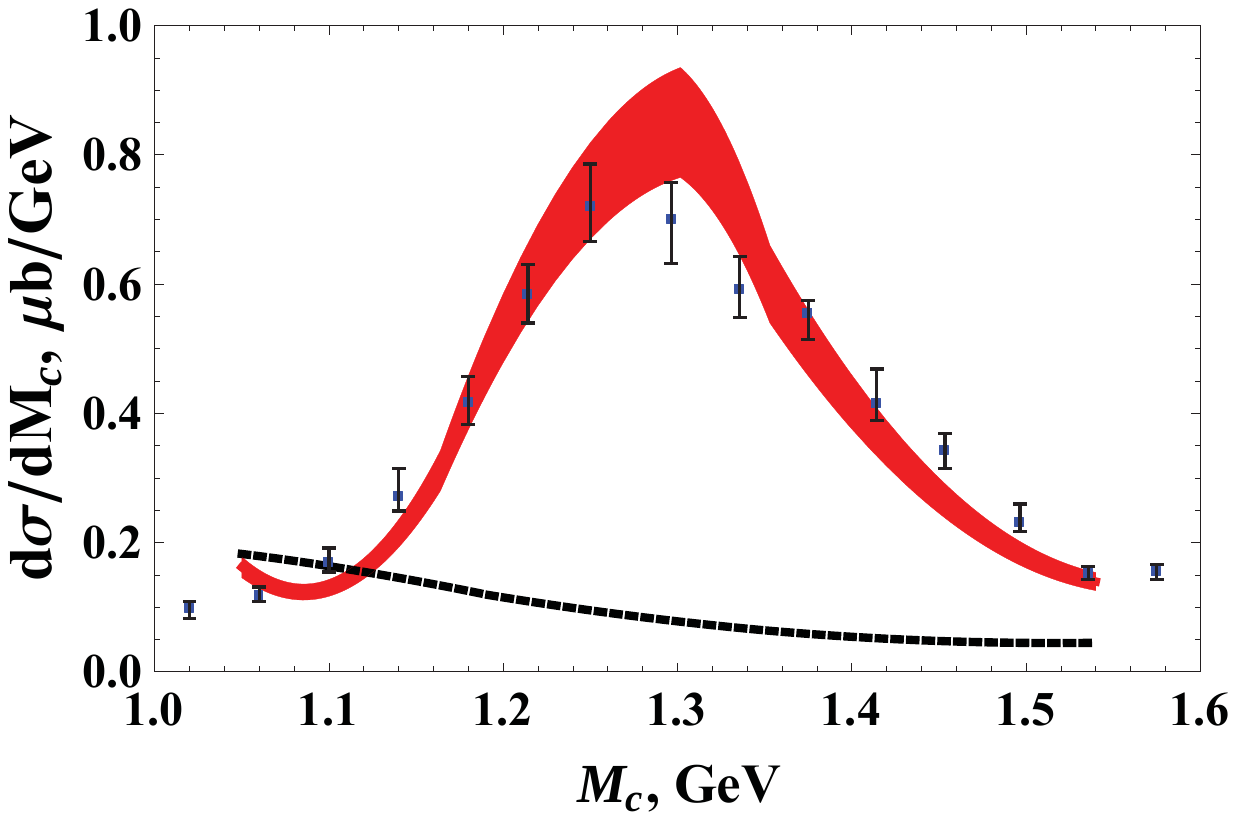}\\
b)\includegraphics[width=0.35\textwidth]{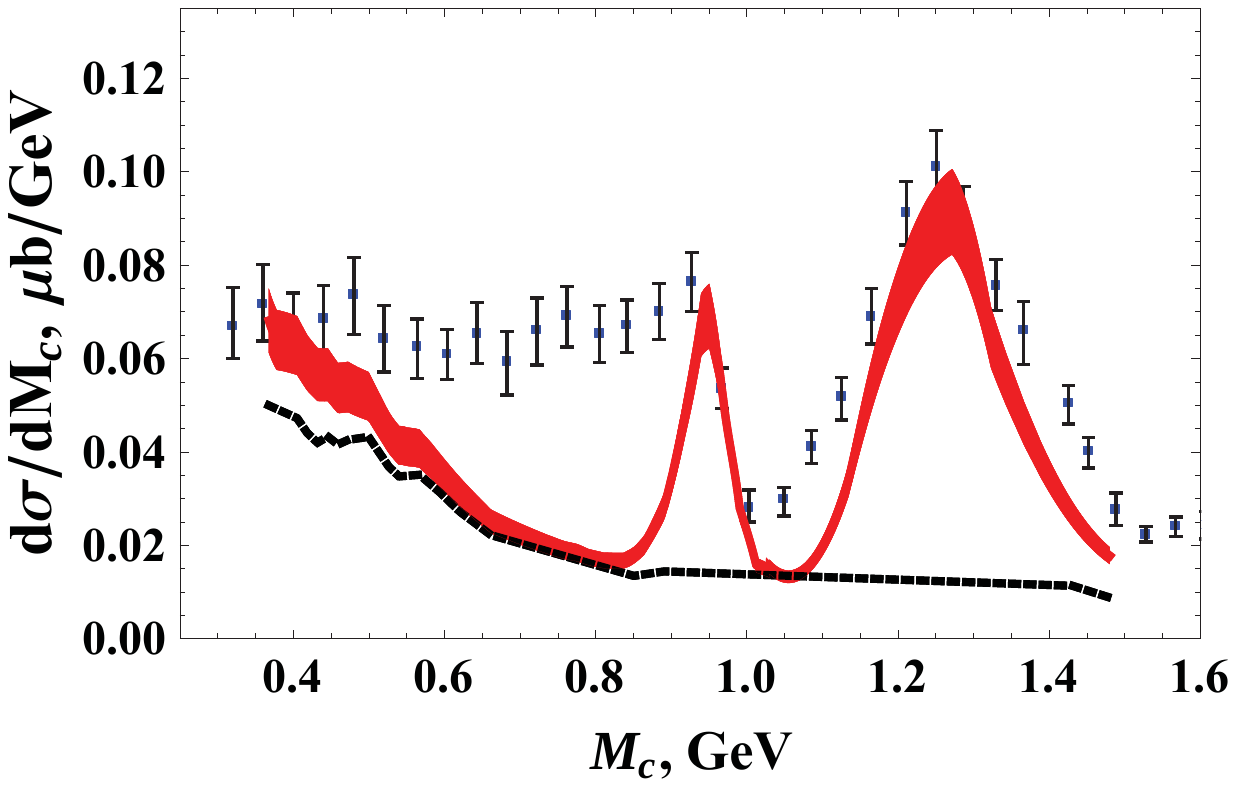}
\caption{\label{fig:CDF} The data on the process $p+\bar{p}\to p+\pi^++\pi^-+\bar{p}$ (CDF collaboration~\cite{CDFdata1},\cite{CDFdata2}) at $\sqrt{s}=1.96$~TeV:
(a)  $|\eta_{\pi}|<1.3$, $|y_{\pi\pi}|<1$, $p_{T,\pi}>0.4$~GeV; 
(b) $|\eta_{\pi}|<1.3$, $|y_{\pi\pi}|<1$, $p_{T,\pi}>0.4$~GeV, $p_{T, \pi\pi}>1$~GeV.  Curves correspond to $\Lambda_{\pi}=1.2$~GeV in the off-shell pion form factor~(\ref{eq:offshellFpi}) and couplings from~(\ref{PPfcouplings}).
Solid upper curves correspond to the sum of all amplitudes and dashed lower ones represent the continuum contribution. Fluctuations in solid curves are due to complex Monte-Carlo integration process, and they can be smoothed by increasing the integration accuracy. Thickness of the solid curves corresponds to the errors of Monte-Carlo calculations.
}
\end{center}
\end{figure}   

Let us look at the ISR~\cite{ISRdata1},\cite{ISRdata2} and CDF~\cite{CDFdata1},\cite{CDFdata2} data with 
parameter $\Lambda_{\pi}$, which we use to describe the data from STAR
collaboration. Different cases are depicted on Figs.~\ref{fig:ISR}-\ref{fig:CDF}.

We see strong underestimation of the ISR data (by factor of $3$). For 
these low energies we have to take into account possible corrections
to pion-proton amplitudes (contributions from secondary 
reggeons are strong), since our approach describe data well
only for energies greater than $\sim 3$~GeV. And in each 
shoulder ($T_{\pi p}$ amplitude in Fig.~\ref{fig3:MY4CASES} (a),(b)) energy can be less than $3$~GeV. Also contributions
from single and double dissociations are possible.

As to the CDF data in the Fig.~\ref{fig:CDF}(a), it is close to the predictions at the region of $f_2(1270)$ meson. And another part of the CDF data in 
the Fig.~\ref{fig:CDF}(b) is close to the prediction, but the curve does not fit the
data at the region of $f_2(1270)$ meson. It also looks strange and has no explanation
at the moment, since
this is the data from the same experiment. This is may be due to interference effects with $\gamma\gamma$ or $\gamma\mathbb{O}$ fusion in the central production process and corrections to $\pi\pi$ final interaction.

\begin{figure}[hbt!]
\begin{center}
a)\includegraphics[width=0.35\textwidth]{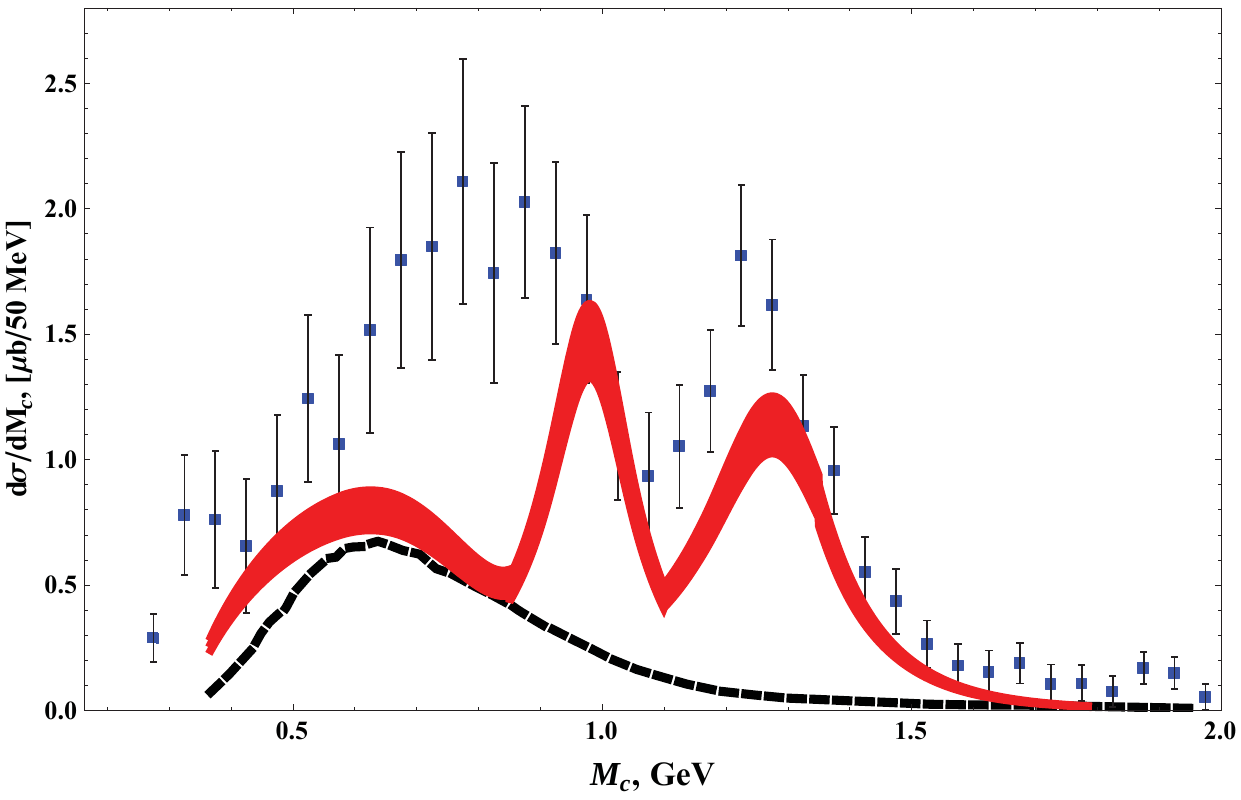}\\
b)\includegraphics[width=0.35\textwidth]{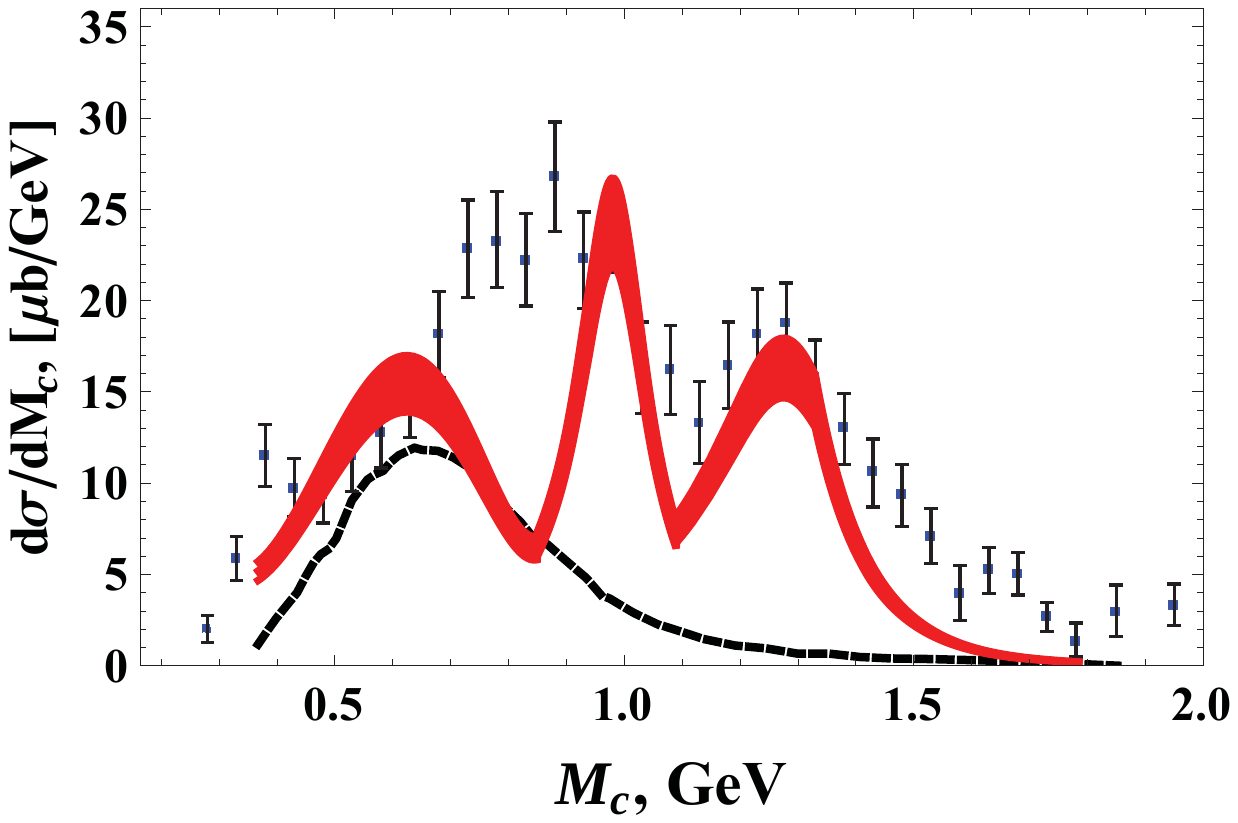}\\
c)\includegraphics[width=0.35\textwidth]{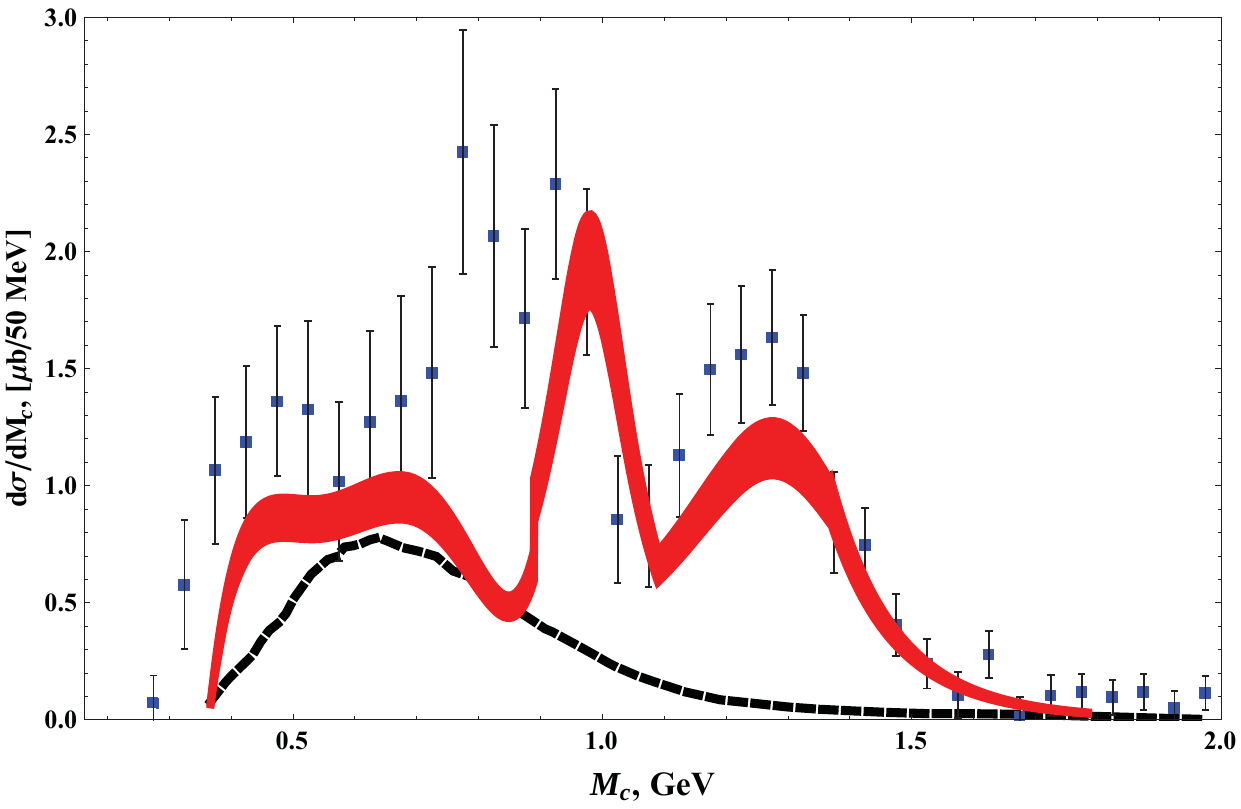}\\
\caption{\label{fig:CMS} The data on the process $p+p\to p+\pi^++\pi^-+p$ (CMS collaboration~\cite{CMSdata1},\cite{CMSdata2}): 
(a) at $\sqrt{s}=5$~TeV with cuts $|\eta_{\pi}|<2.4$, $p_{T,\pi}>0.2$~GeV; 
(b) at $\sqrt{s}=7$~TeV with cuts $|y_{\pi}|<2$, $p_{T,\pi}>0.2$~GeV;
(c) at $\sqrt{s}=13$~TeV with cuts $|\eta_{\pi}|<2.4$, $p_{T,\pi}>0.2$~GeV. Curves correspond to $\Lambda_{\pi}=1.2$~GeV in the off-shell pion form factor~(\ref{eq:offshellFpi}) and couplings from~(\ref{PPfcouplings}).
Solid upper curves correspond to the sum of all amplitudes and dashed lower ones represent the continuum contribution. Thickness of the solid curves corresponds to the errors of Monte-Carlo calculations.
}
\end{center}
\end{figure}   

\subsection{CMS data and predictions}

In Fig.~\ref{fig:CMS} one can see the recent data from the CMS collaboration
and curves of our model, which correspond to $\Lambda_{\pi}=1.2$~GeV in the off-shell pion form factor~(\ref{eq:offshellFpi}) and couplings~(\ref{PPfcouplings}) from the fit to the STAR data on Fig.~\ref{fig:starNEW}. Here our predictions are very close to the data except for the region
$M_c\sim 0.8\pm 0.1$~GeV, as in all datasets.

\section*{Summary and conclusions}

In this paper we have considered the process LM CEDP of di-pions and its description in the framework of the Regge-eikonal approach. 
Here we summarize all the facts and conclusions:
\begin{itemize}
\item The result is crucially dependent on the choice of $\Lambda_{\pi}$ in the off-shell pion form factor, i.e.
on $\hat{t}$ (virtuality of the pion) dependence. In the present approach the best description of the old STAR data~\cite{STARdata1},\cite{STARdata2} is given by the case with $\Lambda_{\pi}=1.2$~GeV and couplings:
\begin{eqnarray}
g_{\mathbb{PP}f_0(500)} &=& 0.88,\nonumber\\
g_{\mathbb{PP}f_0(980)} &=& 0.43,\nonumber\\
g_{\mathbb{PP}f_2(1270)} &=& 1.72.
\label{PPfcouplings}
\end{eqnarray}
The couplings of Pomeron to $f_{0,2}$ can be compared to the value $0.64$, which is 
obtained in~\cite{GodizovResonances}.

\item the model shows that contributions from $\rho$ meson are not so significant at available energies, but we have a dip in the theoretical curves at the region of $\rho$ production (at CMS energies, for example, it is obvious). This also should be explained in further investigations.

\item Rescattering corrections (``soft survival probability'') in $pp$ and $\pi p$ make a significant contribution to the values and form of distributions.

\item If we try to fit the data from STAR~\cite{STARdata3}-\cite{STARdata5}, we find that the curves with parameters fixed from~\cite{STARdata1},\cite{STARdata2} underestimate the data in the region of $\rho$ meson
as in the Fig.\ref{fig:starOLD}(a). The data are close to predictions only in the regions of $f$ mesons. These effects have no explanation at the moment,
and may be due to some complex dependance of Pomeron-Pomeron-meson 
couplings on $t_{1,2}$ and $\phi_0$, wrong normalization of the data or
missed contributions from some other processes like low mass difractive dissociation.

\item We have strong underestimation of the ISR data (by factor of $3$)~\cite{ISRdata1},\cite{ISRdata2}  and contradictory description of 
two parts of the CDF data~\cite{CDFdata1},\cite{CDFdata2} with the same parameters (see Figs.\ref{fig:ISR},\ref{fig:CDF}).

\item However, the CMS data at all energies are described rather well (see Fig.~\ref{fig:CMS}). Discrepancy is observed only in the region $M_c\sim 0.8\pm 0.2$~GeV as for other energies.
\end{itemize}

The main open problems regarding model parameters are related 
to interference terms (we have to know all cut-off form-factor parameters, 
couplings of pions and reggeons to resonances and 
their dependance on $t_{1,2}$ and $\phi_0$). To fix the 
model parameters correctly we need the comparison
with precise (exclusive) experimental data (STAR, ATLAS+ALFA, CMS+TOTEM)
simultaneously in several differential observables,
e.g. the differential distributions $d\sigma/dt_1$, $d\sigma/d\phi_0$,
the angular distributions in the $\pi^+\pi^-$ rest system and others. It should be done in further works.

We have to take into accout also effects
like the interference with $\gamma\gamma\to\pi\pi$ and $\gamma\mathbb{O}\to\pi\pi$ processes, effects related to the irrelevance and possible modifications of the 
Regge approach (for the virtual pion exchange) in this kinematical region, as was discussed in the introduction, corrections to 
pion-pion scattering at low $M_{\pi\pi}$, corrections to
$T_{\pi p}(s,t)$ for $\sqrt{s}<3$~GeV, difference of resonance peaks from the Breit-Wigner functions. In further works we will take into account possible modifications of the model for best description of the data. 

This model will be implemented to the Monte-carlo event generator ExDiff~\cite{ExDiffmanual}. It is possible to 
calculate LM CEDP for other di-hadron final states ($p\bar{p}$ for ``Odderon'' hunting, $K^+K^-$, $\eta\eta^{\prime}$ and 
so on), which are also very informative for our understanding of diffractive mechanisms in strong interactions.


\section*{Acknowledgements}

I am grateful to Vladimir Petrov, Anton Godizov and Piotr Lebiedowicz for useful discussions and help.  

\section*{Appendix A. Kinematics of LM CEDP}
\label{app:kinematics}

\begin{figure}[hbt!]
	\begin{center}
		\includegraphics[width=0.45\textwidth]{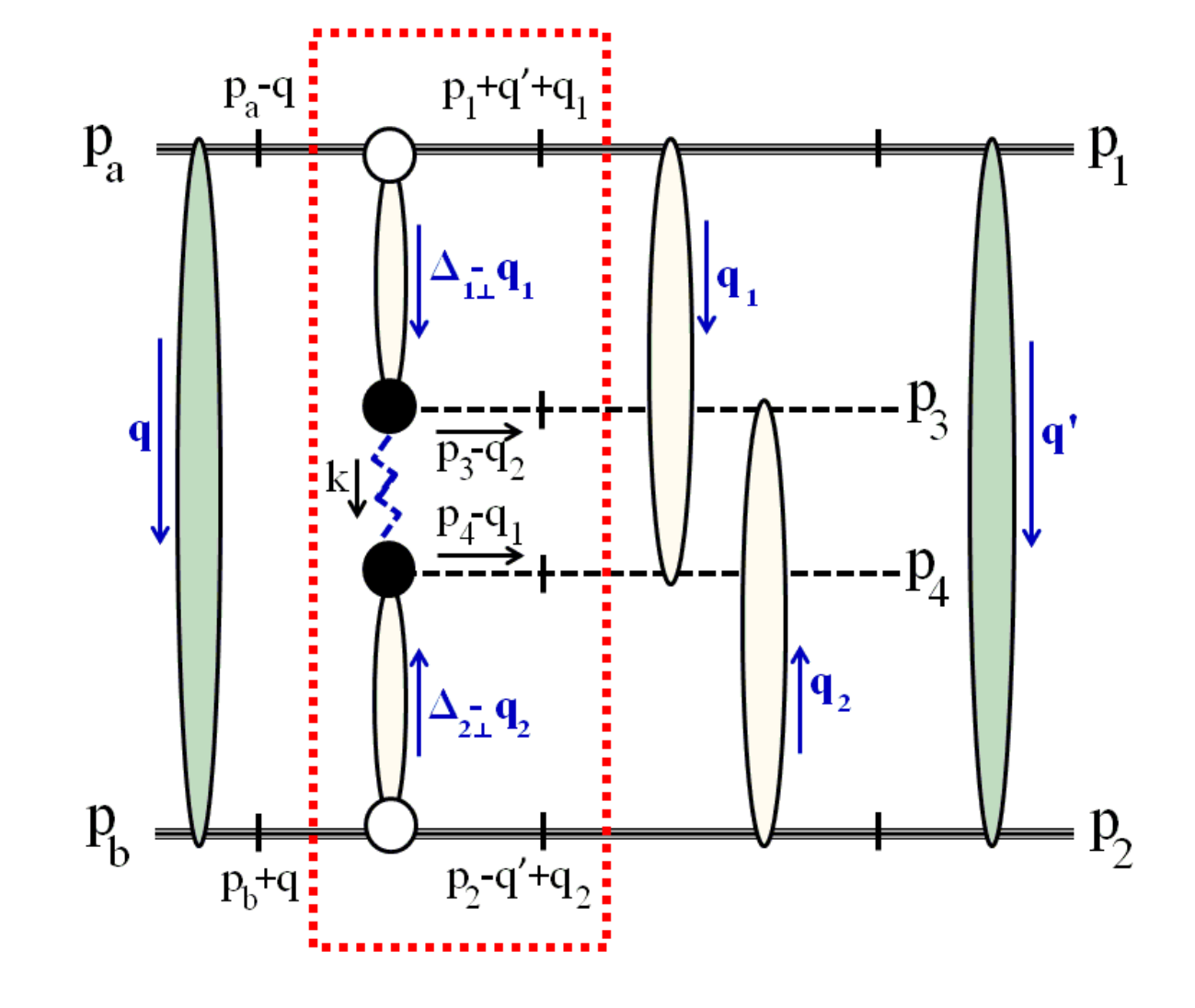}
		\caption{\label{fig4:totalkin} Total amplitude of the process of double pion LM CEDP $p+p\to p+\pi^++\pi^-+p$ with detailed kinematics. Proton-proton rescatterings in the initial and final states are depicted as black blobes, and pion-proton subamplitudes are also shown as shaded blobes. All momenta are shown. Basic part of the amplitude, $M_0$ (see eq.~(\ref{eq:M0})), without corrections is circled by a dotted line. Crossed lines are on mass shell. Here $\Delta_{1\perp}=\Delta_1-q-q^{\prime}$, $\Delta_{2\perp}=\Delta_2+q+q^{\prime}$, $\hat{t}=k^2=(\Delta_{1\perp}-q_1-p_3+q_2)^2$, $\hat{u}=(\Delta_{1\perp}-q_1-p_4)^2$, $\hat{s}=(p_3+p_4-q_1-q_2)^2$.}
	\end{center}
\end{figure}

 The $2\to 4$ process $p(p_a)+p(p_b)\to p(p_1)+\pi(p_3)+\pi(p_4)+p(p_2)$ can be
described as follows (the notation for any momentum is $k=(k_0,k_z;\vec{k})$, $\vec{k}=(k_x,k_y)$):
\begin{eqnarray}
 p_a&=&\left( 
\frac{\sqrt{s}}{2}, \beta\frac{\sqrt{s}}{2}; \vec{0} 
\right),\;
p_b=\left( 
\frac{\sqrt{s}}{2}, -\beta\frac{\sqrt{s}}{2}; \vec{0} 
\right),\nonumber\\
p_{1,2}&=&
\left( 
E_{1,2},p_{1,2z}; \vec{p}_{1,2\perp}
\right),
E_{1,2}=\sqrt{p_{1,2z}^2+\vec{p}_{1,2\perp}^2+m_p^2},\;\nonumber\\
p_{3,4}&=&
\left( 
m_{3,4\perp}\mathrm{ch}\;y_{3,4}, m_{3,4\perp}\mathrm{sh}\;y_{3,4}; \vec{p}_{3,4\perp}
\right) = \nonumber\\
&=&\left( 
\sqrt{m_{\pi}^2+\vec{p}_{3,4\perp}^2\mathrm{ch}^2\eta_{3,4}}, |\vec{p}_{3,4\perp}|\;\mathrm{sh}\;\eta_{3,4}; \vec{p}_{3,4\perp}
\right), \nonumber\\
m_{i\perp}^2&=&m_i^2+\vec{p}_{i\perp}^2,\; m_{1,2}=m_p,\; m_{3,4}=m_{\pi}, \nonumber\\ 
\vec{p}_{4\perp}&=&-\vec{p}_{3\perp}-\vec{p}_{1\perp}-\vec{p}_{2\perp},\;
\nonumber\\
\beta&=&\sqrt{1-\frac{4m_p^2}{s}},\; s=(p_a+p_b)^2,\; s^{\prime}=(p_1+p_2)^2.
\label{eq2:kinmomenta}
\end{eqnarray}
Here $y_i$ ($\eta_i$) are rapidities (pseudorapidities) of final pions.

 Phase space of the process in terms of the above variables is the following
\begin{eqnarray}
 d\mathrm{\Phi}_{2\to 4} &=& 
 \left(2\pi\right)^4\delta^4\left( p_a+p_b-\sum_{i=1}^4p_i\right) \prod_{i=1}^4 \frac{d^3p_i}{(2\pi)^32E_i}=
 \nonumber\\
 &=& \frac{1}{2^4(2\pi)^8}
 \prod_{i=1}^3 p_{i\perp}dp_{i\perp}d\phi_i\cdot dy_3dy_4 \cdot {\cal J};
  \nonumber\\
 {\cal J}&=& \frac{dp_{1z}}{E_1} \frac{dp_{2z}}{E_2} 
 \delta\left( \sqrt{s}-\sum_{i=1}^4 E_i\right)
 \delta\left( \sum_{i=1}^4 p_{iz}\right)=\nonumber\\
 &=&\frac{1}{\left| \tilde{E}_2\tilde{p}_{1z}-\tilde{E}_1\tilde{p}_{2z}\right|},
\label{eq3:kinPhSp}  
\end{eqnarray} 
where $p_{i\perp}=\left| \vec{p}_{i}\right|$, $\tilde{p}_{1,2z}$ are appropriate roots of the system
\begin{equation}
\begin{cases}
\label{eq4:sys} &A=\sqrt{s}-E_3-E_4=\sqrt{m_{1\perp}^2+p_{1z}^2}+\sqrt{m_{2\perp}^2+p_{2z}^2},\\
 &B=-p_{3z}-p_{4z}=p_{1z}+p_{2z},
\end{cases}
\end{equation}
\begin{eqnarray}
 \tilde{p}_{1z}&=& \frac{B}{2}+\frac{1}{2(A^2-B^2)}\left[ 
B\left( m_{1\perp}^2-m_{2\perp}^2\right) + A\cdot\lambda_0^{1/2}
\right],\nonumber\\
\label{eq4:sysroot} \lambda_0&=&\lambda\left( A^2-B^2,m_{1\perp}^2,m_{2\perp}^2\right). 
\end{eqnarray}
Here $\lambda(x,y,z)=x^2+y^2+z^2-2xy-2xz-2yz$, and then ${\cal J}^{-1}=\lambda_0^{1/2}/2$.

For the differential cross-section we have
\begin{eqnarray}
\frac{d\sigma_{2\to 4}}{\prod_{i=1}^{3\phantom{I}} dp_{i\perp}d\phi_i\cdot dy_3dy_4 }&=&
\frac{1}{2\beta s}\cdot\frac{\prod_{i=1}^3 p_{i\perp}}{2^4(2\pi)^8\cdot\frac{1}{2}\lambda_0^{1/2}}
\left| T\right|^2=
\nonumber\\
\label{eq5:dcsdall} &=& \frac{\prod_{i=1}^3 p_{i\perp}}{2^{12}\pi^8\beta s \lambda_0^{1/2}}\left| T \right|^2.
\end{eqnarray}
Pseudorapidity is more convenient experimental variable, and we can use the transform
\begin{equation}
\frac{dy_i}{d\eta_i}=\frac{p_{i\perp}\mathrm{ch}\eta_i}{\sqrt{m_i^2+p_{i\perp}^2\mathrm{ch}^2\eta_i}}
\end{equation}
to get the differential cross-section in pseudorapidities.

In some cases it is convenient to use other variables for the integration of the cross-section and
calculation of distributions on central mass. For these cases we have:
\begin{eqnarray}
 d\mathrm{\Phi}_{2\to 4}  &=& \frac{1}{2^4(2\pi)^8}
 \prod_{i=1}^2 dt_i d\phi_i\cdot M_c dM_c d\eta_c dc^* d\phi^*\cdot {\cal J}^{\prime};
  \nonumber\\
 {\cal J}^{\prime}&=& \frac{\beta_M}{4\beta_1\beta_2 s} \frac{dy_c}{d\eta_c},\, \,\,\,\,
 \beta_i \simeq \sqrt{1+\frac{4(m_p^2-(1-\xi_i)t_i)}{\beta^2s^2(1-\xi_i)^2}}; \nonumber\\
\beta_M &=& \sqrt{1-\frac{4m_{\pi}^2}{M_c^2}},\, \,
\frac{dy_c}{d\eta_c}=\frac{p_{c\perp}\mathrm{ch}\eta_c}{\sqrt{M_c^2+p_{c\perp}^2\mathrm{ch}^2\eta_c}};
\label{eq3:kinPhSpMc}  
\end{eqnarray} 

\begin{eqnarray}
\frac{d\sigma_{2\to 4}}{\prod_{i=1}^{2\phantom{I}} dt_{i}d\phi_i dM_c d\eta_c dc^* d\phi^*}&=&
\frac{1}{2\beta s}\cdot\frac{M_c \beta_M \frac{dy_c}{d\eta_c}}{2^4(2\pi)^8\cdot 4\beta_1\beta_2 s}
\left| T\right|^2
\nonumber\\
\label{eq5:dcsdallMc} &=& \frac{M_c \beta_M \frac{dy_c}{d\eta_c}}{2^{15}\pi^8\beta\beta_1\beta_2 s^2 }\left| T \right|^2,
\end{eqnarray}
where $c^*=\cos\theta^*$, $\theta^*$ and $\phi^*$ are polar and azimuthal angles of the pion momenta
in the $\pi^+\pi^-$ rest frame, $M_c$ is the di-pion mass, $\eta_c$ is the di-pion 
pseudorapidity, $t_1=(p_a-p_1)^2$, $t_2=(p_b-p_2)^2$ and
$$
\xi_{1,2}\simeq \sqrt{\frac{M_c^2-t_1-t_2+2\sqrt{t_1t_2}\cos(\phi_1-\phi_2)}{s}}
\mbox{\bf e}^{\pm y_c}.
$$

For exact calculations of elastic subprocesses (see Fig.~\ref{fig4:totalkin}) of the type\\ $a(p_1)+b(p_2)\to c(p_1-q_{el})+d(p_2+q_{el})$:
\begin{eqnarray}
q_{el}&=&\left( q_0,q_z; \vec{q}\right),\nonumber\\
q_z&=&-\frac{b}{2a}\left( 1-\sqrt{1-\frac{4ac}{b^2}}\right),\nonumber\\ q_0&=&\frac{A_0q_z+\vec{p}_{1\perp}\vec{q}+\vec{p}_{2\perp}\vec{q}}{A_z},\nonumber\\
a&=&A_z^2-A_0^2,\; b=-2
\left( 
A_z\cdot {\cal D}+A_0 \left( \vec{p}_{1\perp}\vec{q}+\vec{p}_{2\perp}\vec{q}\right)
\right),\nonumber\\
c&=&2 A_z B_z-\left( \vec{p}_{1\perp}\vec{q}+\vec{p}_{2\perp}\vec{q}\right)^2+\vec{q}^2A_z^2,\nonumber\\
A_0&=&p_{1z}+p_{2z},\; A_z=p_{10}+p_{20}, \nonumber\\
B_0&=&p_{1z}\cdot\vec{p}_{2\perp}\vec{q}-p_{2z}\cdot\vec{p}_{1\perp}\vec{q},\nonumber\\
B_z&=&p_{10}\cdot\vec{p}_{2\perp}\vec{q}-p_{20}\cdot\vec{p}_{1\perp}\vec{q},\nonumber\\
{\cal D}&=&p_{1z}p_{20}-p_{2z}p_{10},\label{eq6:qel}
\end{eqnarray}
and $q_{el}^2\simeq -\vec{q}^2$.

\section*{Appendix B. Regge-eikonal model for elastic pro\-ton-pro\-ton and pion-pro\-ton scattering}
\label{app:godizovmodel}

Here is a short review of formulae for the Regge-eikonal approach~\cite{godizovpp}, \cite{godizovpip}, which we use to estimate rescattering corrections in the proton proton and pion proton channels.

Amplitudes of elastic proton-proton and pion-proton scattering are expressed in terms of eikonal functions
\begin{eqnarray}
T_{pp,\pi p}^{el}(s,b)&=&\frac{\mathrm{e}^{-2\Omega_{pp,\pi p}^{el}(s,b)}-1}{2\mathrm{i}},\nonumber\\
\Omega_{pp, \pi p}^{el}(s,b) &=& -\mathrm{i}\,\delta_{pp,\pi p}^{el}(s,b),\nonumber\\
\delta^{el}_{pp,\pi p}(s,b)&=&\frac{1}{16\pi s}\int_0^\infty d(-t) J_0(b\sqrt{-t}) \delta^{el}_{pp,\pi p}(s,t).
\label{eq:elamplitudes}
\end{eqnarray}


\begin{eqnarray}
&& \delta^{el}_{pp}(s,t)\simeq\nonumber\\
&& g_{pp\mathbb{P}}(t)^2\left(
\mathrm{i}+\tan\frac{\pi(\alpha_{\mathbb{P}}(t)-1)}{2}) 
\right) \pi \alpha^{\prime}_{\mathbb{P}}(t)
\left( \frac{s}{2s_0}\right)^{\alpha_{\mathbb{P}}(t)},\nonumber\\
&& \alpha_{\mathbb{P}}(t)=1+\frac{\alpha_{\mathbb{P}}(0)-1}{1-\frac{t}{\tau_a}}\,,\, g_{pp\mathbb{P}}(t)=\frac{g_{pp\mathbb{P}}(0)}{\left( 1-a_g t\right)^2}.
\label{eq:godizovpp}
\end{eqnarray}

\begin{eqnarray}
&& \delta^{el}_{\pi p}(s,t)\simeq\nonumber\\
&& \left(
\mathrm{i}+\tan\frac{\pi(\alpha_{\mathbb{P}}(t)-1)}{2}) 
\right) \beta_{\mathbb{P}}(t)
\left( \frac{s}{s_0}\right)^{\alpha_{\mathbb{P}}(t)},\nonumber\\
&& +\left(
\mathrm{i}+\tan\frac{\pi(\alpha_{f}(t)-1)}{2}) 
\right) \beta_{f}(t)
\left( \frac{s}{s_0}\right)^{\alpha_{f}(t)},
\label{eq:godizovpip1}
\end{eqnarray}

\begin{eqnarray}
&& \alpha_{\mathbb{P}}(t)=1+p_1\left[ 
1-p_2 t\left( 
\arctan\left( p_3-p_2t\right) 
-\frac{\pi}{2}
\right)
\right],\nonumber\\
&& \alpha_f(t)=\left(
\frac{8}{3\pi}\gamma(\sqrt{-t+c_f})
\right)^{1/2},\nonumber\\
&& \gamma(\mu)=\frac{4\pi}{11-\frac{2}{3}n_f}\left( 
\frac{1}{\ln\frac{\mu^2}{\Lambda^2}}+\frac{1}{1-\frac{\mu^2}{\Lambda^2}}
\right),\nonumber\\
&& \beta_{\mathbb{P}}(t)=B_{\mathbb{P}}\mathrm{e}^{b_{\mathbb{P}}t}(1+d_1t+d_2t^2+d_3 t^3+d_4 t^4),\nonumber\\
&& \beta_f(t)=B_f \mathrm{e}^{b_f t}.
\label{eq:godizovpip2}
\end{eqnarray}
Parameters can be found in tables~\ref{tab1} and~\ref{tab2}.

\begin{table}[ht!]
	\caption{\label{tab1} Parameters for proton-proton elastic scattering amplitude.}
	\begin{center}
		\begin{tabular}{|c|c|}
			\hline
			Parameter &   Value  \\	
			\hline	
			$\alpha_{\mathbb{P}}(0)-1$ & $0.109$  \\	           	
			\hline
	$\tau_a$ &  $0.535$~GeV$^2$  \\	           	
	\hline	
	$g_{pp\mathbb{P}}(0)$ & $13.8$~GeV  \\	           	
	\hline				
	$a_g$ &    $ 0.23$~GeV$^{-2}$  \\	           	
	\hline					
		\end{tabular}
	\end{center}
\end{table}  

\begin{table}[ht!]
	\caption{\label{tab2} Parameters for pion-proton elastic scattering amplitude.}
	\begin{center}
		\begin{tabular}{|c|c|}
			\hline
			Parameter &   Value  \\	
			\hline	
			$B_{\mathbb{P}}$ & $26.7$  \\	           	
			\hline
			$b_{\mathbb{P}}$ &  $2.36$~GeV$^{-2}$  \\	           	
			\hline	
			$d_1$ & $0.38$~GeV$^{-2}$  \\	           	
			\hline				
			$d_2$ &    $0.3$~GeV$^{-4}$  \\	           	
			\hline	
			$d_3$ &    $-0.078$~GeV$^{-6}$  \\	           	
			\hline
			$d_4$ &    $0.04$~GeV$^{-8}$  \\	           	
			\hline
			$B_f$ &    $67$  \\	           	
			\hline
			$b_f$ &    $1.88$~GeV$^{-2}$  \\	           	
			\hline														
		\end{tabular}
	\end{center}
\end{table} 

\begin{eqnarray}
V_{pp}(s,q^2)&=&\int d^2\vec{b} \,\mathrm{e}^{\mathrm{i}\vec{q}\vec{b}}
\sqrt{1+2\mathrm{i}T_{pp}^{el}(s,b)}=\nonumber\\
&=& \int d^2\vec{b}\, \mathrm{e}^{\mathrm{i}\vec{q}\vec{b}} 
\mathrm{e}^{-\Omega_{pp}^{el}(s,b)}=\nonumber\\
&=& (2\pi)^2\delta^2\left( \vec{q}\right) + 2\pi \bar{T}_{pp}(s,q^2),
\label{eq:Vpp1}\\
\bar{T}_{pp}(s,q^2)&=& \int_0^\infty b\,db\, J_0\left( b\sqrt{-q^2}\right)
\left[ 
\mathrm{e}^{-\Omega_{pp}^{el}(s,b)}-1
\right]
\label{eq:Vpp2}\\
S_{\pi p}(s,q^2)&=&\int d^2\vec{b} \,\mathrm{e}^{\mathrm{i}\vec{q}\vec{b}}
\left(
1+2\mathrm{i}T_{\pi p}^{el}(s,b)
\right)=\nonumber\\
&=& \int d^2\vec{b}\, \mathrm{e}^{\mathrm{i}\vec{q}\vec{b}} 
\mathrm{e}^{-2\Omega_{\pi p}^{el}(s,b)}=\nonumber\\
&=& (2\pi)^2\delta^2\left( \vec{q}\right) + 2\pi \bar{T}_{\pi p}(s,q^2),
\label{eq:Vpip1}\\
\bar{T}_{\pi p}(s,q^2)&=& \int_0^\infty b\,db\, J_0\left( b\sqrt{-q^2}\right)
\left[ 
\mathrm{e}^{-2\Omega_{\pi p}^{el}(s,b)}-1
\right]
\label{eq:Vpip2}
\end{eqnarray}

Here we take $S_{\pi p}(s,t)=S_{\pi^+ p}(s,t)=S_{\pi^- p}(s,t)$ and
\begin{equation}
T^{el}_{\pi^+p}(s,t)=T^{el}_{\pi^-p}(s,t)=4\pi s \bar{T}_{\pi p}(s,t)
\label{eq:Tpipexact}
\end{equation}

Approach~(\ref{eq:godizovpip1}) describes the data on pion-proton scattering better even at low energies, that
is why we use it instead of the one presented in~\cite{godizovpp}.

Functions $\bar{T}_{pp}$ and $\bar{T}_{\pi p}$ are convenient for numerical
calculations, since its oscillations are not so strong.

\section*{Appendix C. Primary amplitudes for CEDP and ERVM resonance production.}
\label{app:ampres}

Here is the short review of the formulae, which can be obtained by the use of Refs.~\cite{godizovmesons}, \cite{GodizovResonances} and \cite{CEDPLIBbasis}-\cite{CEDPLIBrho}.

Let us introduce the general diffractive factor
\begin{equation}
F_{\mathbb{P}}(t,\xi) = g_{pp\mathbb{P}}(t)^2\left(
\mathrm{i}+\tan\frac{\pi(\alpha_{\mathbb{P}}(t)-1)}{2}) 
\right) \frac{\pi \alpha^{\prime}_{\mathbb{P}}(t)}{\xi^{\alpha_{\mathbb{P}}}(t)}.
\label{eq:Pomerongenfactor}
\end{equation}

\subsection*{\bf $f_0$ production}
For $f_0$ production we have the following expression
\begin{eqnarray}
\phantom{M_{000}}&&\!\!\!\!\!M_0^{pp\to p \{f_0\to\pi^+\pi^- \} p} = \nonumber\\
&=& -F_{\mathbb{P}}(t_1,\xi_1) F_{\mathbb{P}}(t_2,\xi_2) \, g_{\mathbb{P}\mathbb{P}f_0}(t_1,t_2,M_c^2)\times
\nonumber\\
&\times&  
\frac{g_{f_0\pi\pi} \left({\cal F}(M_c^2,m_{f_0}^2)\right)^2F_M(t_1)F_M(t_2)}{(M_c^2-m_{f_0}^2)+B_{f_0}(M_c^2,m_{f_0}^2)},
\label{eq:AMPf0GEN}
\end{eqnarray}
where ($M_c > 2m_{\pi}$)
\begin{eqnarray}
&& \!\!B_{f_0}(M_c^2,m_{f_0}^2) =\nonumber\\
&&= \mathrm{i}\; \Gamma_{f_0}
\left({\cal F}(M_c^2)\right)^2\left[\frac{1-4m_{\pi}^2/M_c^2}{1-4m_{\pi}^2/m_{f_0}^2}\right]^{1/2} \label{eq:Bf0BW}\\
&& \!\!\!\! {\cal F}(M_c^2,m_f^2) = F^{\mathbb{P}\mathbb{P}f}(M_c^2,m_f^2)=F^{f\pi\pi}(M_c^2,m_f^2)=\nonumber\\
&& \!\!\!\! 
=\exp\left( \frac{-(M_c^2-m_{f}^2)^2}{\Lambda_{f}^4}\right),\, \Lambda_{f}\sim 1\,\mathrm{GeV},
\label{eq:FFf0}\\
&& \!\!\!\! F_M(t) = 1/(1-t/m_0^2),\, m_0^2 = 0.5\,\mathrm{GeV}^2
\label{eq:FMLib}
\end{eqnarray}
${\cal F}(M_c^2,m_f^2)$ and $F_M(t)$ are off-shell phenomenological form-factors introduced in~\cite{CEDPLIBbasis}-\cite{CEDPLIBrho} to make more good description of the data. Here we fix mass and width of $f_0(500)$ and $f_0(980)$ mesons as in~\cite{CEDPLIBf0f2}
\begin{eqnarray}
m_{f_0(500)} &=& 0.6\,\mathrm{GeV},\, \Gamma_{f_0(500)} = 0.5\,\mathrm{GeV}\label{eq:f0500pars},\\
m_{f_0(980)} &=& 0.98\,\mathrm{GeV},\, \Gamma_{f_0(980)} = 0.07\,\mathrm{GeV}\label{eq:f0500pars},
\end{eqnarray}
and also their couplings to pions
\begin{equation}
g_{f_0(500)\pi\pi} = 3.37,\, g_{f_0(500)\pi\pi} = 1.55,
\label{eq:f0500gpipi}
\end{equation}
assuming also $\Gamma(f_0\to\pi\pi)/\Gamma_{f_0}=100\%$.

In this work coupling of Pomeron to meson $g_{\mathbb{P}\mathbb{P}f_0}$ 
is the constant and can be extracted from the experimental data as a 
free parameter. In~\cite{GodizovResonances} this coupling is proposed to be $0.64$~GeV for all mesons
related to the ``glueball state''. Here due to simplifications these couplings are of
the same order, but may differ by factor $2\div 3$ (see the main text).

\subsection*{\bf $f_2$ production}
For $f_2(1270)$ production we have to modify the expression~(\ref{eq:AMPf0GEN})
due to the tensor nature of this meson. Finally we have
\begin{eqnarray}
\phantom{M}&&\!\!\!\!\!M_0^{pp\to p \{f_2\to\pi^+\pi^- \} p} = \nonumber\\
&=& F_{\mathbb{P}}(t_1,\xi_1) F_{\mathbb{P}}(t_2,\xi_2) \,g_{\mathbb{P}\mathbb{P}f_2}\times
\nonumber\\
&\times&  
\frac{(g_{f_2\pi\pi}/2) \left({\cal F}(M_c^2,m_{f_2}^2)\right)^2F_M(t_1)F_M(t_2)}{(M_c^2-m_{f_2}^2)+B_{f_2}(M_c^2,m_{f_2}^2)} {\cal P}_2,
\label{eq:AMPf2GEN}
\end{eqnarray}
where
\begin{equation}
{\cal P}_2 = (\Delta_1\Delta_{34})^2-\frac{(M_c^2-4m_{\pi}^2)\lambda(M_c^2,t_1,t_2)}{12 M_c^2}
\label{eq:f2Tensorfun}
\end{equation}
is the additional function which can be obtained by contraction of Pomeron-Pomeron-meson vertex with pion-pion-meson one, $\Delta_1=p_a-p_1$, $\Delta_{34}=p_3-p_4$. 
\begin{eqnarray}
&& \!\!\!\!\!\!\!B_{f_2}(M_c^2,m_{f_2}^2) = \nonumber\\
&=&\mathrm{i}\; \Gamma_{f_2}
\left({\cal F}(M_c^2,m_{f_2}^2)\right)^2\left[\frac{1-4m_{\pi}^2/M_c^2}{1-4m_{\pi}^2/m_{f_0}^2}\right]^{5/2} \frac{M_c^4}{m_{f_2}^4}\label{eq:Bf2BW}
\end{eqnarray}
Fixed parameters for $f_2(1270)$ are
\begin{eqnarray}
m_{f_2(1270)} &=& 1.275\,\mathrm{GeV},\, \Gamma_{f_2(1270)} = 0.1851\,\mathrm{GeV},\nonumber\\
g_{f_2(1270)\pi\pi} &=& 9.26\,\mathrm{GeV}^{-1},\, \Lambda_{f_2}=1\,\mathrm{GeV}.
\label{eq:f21270pars}
\end{eqnarray}
After the data analysis the ``scalar'' coupling of the $f_2(1270)$ to Pomeron
is found to be of the order of $2$. Since the Pomeron does not behave as a simple
scalar, the structure of the coupling is, of course, more complicated. In the
``exact'' model like in~\cite{CEDPLIBbasis} Pomeron can be a coherent sum of different spins, and have several couplings to $f$ mesons. But in real life we have to obtain the vertex for any spin $J$ and then
to make a continuation to the complex plane in $J$, like in classical Regge 
theory. This is the conceptual question, which was partially discussed in~\cite{myCEDPpipiContinuum}, but we postpone it to our futher theoretical
works. And in the present work we use simplified model for the $f_2$ production. That is why the extracted value of the ``constant'' coupling could differ from the
real set of tensor couplings.

\subsection*{\bf $\rho_0$ production}
For $\rho_0$ situation is a little bit complicated, since we have to take into account
vector dominance and also $\rho-\omega$ mixing. After all contractions the primary amplitude looks as follows
\begin{eqnarray}
&&\!\!\!\!\!\!M_0^{pp\to p \{\rho_0\to\pi^+\pi^- \} p} = \mathrm{i}\;T^{el}_{\rho_0 p}(s_2,t_2)\frac{C_T^{\rho_0}(t_1){\cal P}_{\rho}}{|t_1|} \times\nonumber\\
&\times& 
\frac{(g_{\rho_0\pi\pi}/2) {\cal F}_{\rho}(M_c^2){\cal F}_{\rho}(t_1)}{(M_c^2-m_{\rho_0}^2)+B_{\rho_0}(M_c^2,m_{\rho_0}^2)}+(1\leftrightarrow 2) ,
\label{eq:AMPrhoGEN}
\end{eqnarray}
$$
s_2=(p_3+p_4+p_2)^2=(p_a-p_1+p_b)^2,
$$
$$
s_1=(p_3+p_4+p_1)^2=(p_b-p_2+p_a)^2.
$$ 
Amplitude 
$T^{el}_{\rho_0 p}(s,t) $ can be obtained by the use of the formulae
similar to (\ref{eq:elamplitudes})-(\ref{eq:godizovpip1}) with
\begin{eqnarray}
 \delta^{el}_{\rho_0 p}(s,t)&\simeq& 
g_{pp\mathbb{P}}(t)g_{\rho\rho\mathbb{P}}(t)\left(
\mathrm{i}+\tan\frac{\pi(\alpha_{\mathbb{P}}(t)-1)}{2}) 
\right)\times\nonumber\\
&\times& \pi \alpha^{\prime}_{\mathbb{P}}(t)
\left( \frac{s}{2s_0}\right)^{\alpha_{\mathbb{P}}(t)},\nonumber\\
  g_{\rho\rho\mathbb{P}}(t) &=& g_{\rho\rho\mathbb{P}}(0) = 7.07\,\mathrm{GeV}\,\,\,\mbox{(see~\cite{godizovmesons})};
\label{eq:godizovrhop}
\end{eqnarray}

\begin{eqnarray}
C_T^{\rho_0}(t) &=& \sqrt{\frac{3\Gamma_{\rho\to e^+e^-}}{\alpha_e m_{\rho}}}\frac{m_{\rho}^2}{m_{\rho}^2-t},\\
{\cal F}_{\rho}(p^2) &=& (1+p^2(p^2-m_{\rho}^2)/\Lambda_{\rho}^4)^{-n_{\rho}},\\
F_1(t)&=&\frac{1-\kappa\, t/(4 m_p^2)}{1-t/(4m_p^2)}(1-t/m_D^2)^{-2},
\end{eqnarray}
\begin{eqnarray}
\kappa&=&\mu_p/\mu_N=2.7928,\, m_D^2=0.71\,\mathrm{GeV}^2,\\
&&\!\!\!\!\!\Gamma_{\rho\to e^+e^-}= 7.04\cdot 10^{-6}\,\mathrm{GeV},\\
&& \phantom{\frac{A}{B}}\!\!\!\!\!\!\!\!\!{\cal P}_{\rho} \simeq 2\sqrt{4\pi\alpha_e}F_1(t) 
\left( \vec{\Delta}_{34}^*\vec{p}_a^* \right),
\end{eqnarray}
${\cal P}_{\rho}$ is the factor that is equal to ``scalar proton-proton-photon vertex''
contracted with the vector of $\rho\pi\pi$ vertex, and the scalar product of transverse
vectors $\left( \vec{\Delta}_{34}^*\vec{p}_a^* \right)$ in the $\rho$ meson
rest frame with 
$$
\Delta_1^*=\{(M_c^2+t_1-t_2)/(2M_c),0,0,\lambda^{1/2}/(2M_c)\}
$$ 
can be expressed in terms of four vectors in the central mass frame of
colliding protons: 
\begin{eqnarray}
&& \left( \vec{\Delta}_{34}^*\vec{p}_a^* \right) = -\Delta_{34}p_a
-\Delta_{34,z}^*p_{a,z}^*,\nonumber\\
&& \Delta_{34,z}^* = -\Delta_1\Delta_{34}\frac{2M_c}{\lambda^{1/2}},\nonumber\\
&&  p_{a,z}^* = \left( 
\frac{p_ap_c(M_c^2+t_1-t_2)}{2M_c^2}-\frac{t_1}{2}
\right)\frac{2M_c}{\lambda^{1/2}},\nonumber\\
&& \lambda \equiv \lambda(M_c^2,t_1,t_2),\, p_c=p_3+p_4.
\end{eqnarray}

Other functions and parameters are
\begin{eqnarray}
&& \!\!\!\!\!\!\!B_{\rho_0}(M_c^2,m_{\rho_0}^2) = \nonumber\\
&=&\mathrm{i}\; \Gamma_{\rho_0}
\left({\cal F}_{\rho}(M_c^2)\right)^2\left[\frac{1-4m_{\pi}^2/M_c^2}{1-4m_{\pi}^2/m_{\rho_0}^2}\right]^{3/2} \frac{M_c^2}{m_{\rho_0}^2}\label{eq:BrhoBW}
\end{eqnarray}
\begin{eqnarray}
m_{\rho_0} &=& 0.7737\,\mathrm{GeV},\, \Gamma_{\rho_0} = 0.1462\,\mathrm{GeV},\nonumber\\
g_{\rho_0\pi\pi} &=& 11.51\,\mathrm{GeV},\, \Lambda_{\rho}=1\,\mathrm{GeV},
\, n_{\rho}=0.5.
\label{eq:rhopars}
\end{eqnarray}

Here we take for all off-shell propagators of resonances simple Breit-Wigner
form, but we can use more complicated expressions, which can be found, for 
example, in~\cite{CEDPLIBbasis}-\cite{CEDPLIBrho}.

\begin{eqnarray}
&& M^U\left( \{ p \}\right) = 
\nonumber\\
&&  
=\int\int 
\frac{d^2\vec{q}}{(2\pi)^2}\frac{d^2\vec{q}^{\prime}}{(2\pi)^2}
\frac{d^2\vec{q}_1}{(2\pi)^2}\frac{d^2\vec{q}_2}{(2\pi)^2}
V_{pp}(s,q^2)
V_{pp}(s^{\prime},q^{\prime 2})\nonumber\\
&& \times\; 
\left[
\left[
S_{\bar{h}p}(\tilde{s}_{14},q_1^2)
M_0^{C}\left( \{\tilde{p}\} \right)
S_{hp}(\tilde{s}_{23},q_2^2)+
(3\leftrightarrow 4)
\right]
+M_0^R\left( \{\tilde{p}\} \right)
\right]  \nonumber\\
&&
\approx\int\int 
\frac{d^2\vec{q}}{(2\pi)^2}
\frac{d^2\vec{q}_1}{(2\pi)^2}\frac{d^2\vec{q}_2}{(2\pi)^2}
S_{pp}(s,q^2) \nonumber\\
&& \times\; 
\left[
\left[
S_{\bar{h}p}(\tilde{s}_{14},q_1^2)
M_0^{C}\left( \{\tilde{p}\} \right)
S_{hp}(\tilde{s}_{23},q_2^2)+
(3\leftrightarrow 4)
\right]
+M_0^R\left( \{\tilde{p}\} \right)
\right]_{q^{\prime}\to 0} \nonumber\\
&& M_0^{C}\left( \{ p \}\right)=
T^{el}_{hp}(s_{13},t_1)
{\cal P}_{h}(\hat{s},\hat{t})
\left[ 
\hat{F}_{h}\left( \hat{t}\right)
\right]^2
T^{el}_{\bar{h}p}(s_{24},t_2), \nonumber
\end{eqnarray}
where functions are defined in~(\ref{eq:Vpp1})-(\ref{eq:Tpipexact}) of Appendix~B, and sets of vectors are
\begin{eqnarray}
&&\{ p \}\equiv \{ p_a,p_b,p_1,p_2,p_3,p_4\}\label{eq:setp}\nonumber\\
&&\{ \tilde{p}\}\equiv \{ p_a-q,p_b+q; p_1+q^{\prime}+q_1,\nonumber\\
&& \;\;\;\;\;\;\;\;\;\;\;\; p_2-q^{\prime}+q_2,p_3-q_2,p_4-q_1 \},\nonumber
\end{eqnarray}
and
\begin{eqnarray}
\tilde{s}_{14}&=&\left( p_1+p_4+q^{\prime} \right)^2,\;
\tilde{s}_{23}=\left( p_2+p_3-q^{\prime} \right)^2,
\label{eq:invarstild}\\
s_{ij}&=&\left( p_i+p_j \right)^2,\;
t_{1,2}=\left( p_{a,b}-p_{1,2} \right)^2,
\label{eq:invars}\\
\hat{s}&=&\left( p_3+p_4 \right)^2,\;
\hat{t}=\left( p_a-p_1-p_3 \right)^2
\label{eq:invarsh}
\end{eqnarray}

\begin{eqnarray}
&& S_{h_1h_2}(s,q^2)=\int d^2\vec{b} \,\mathrm{e}^{\mathrm{i}\vec{q}\vec{b}}
\left(
1+2\mathrm{i}T_{h_1h_2}^{el}(s,b)
\right)=\nonumber\\
&& = \int d^2\vec{b}\, \mathrm{e}^{\mathrm{i}\vec{q}\vec{b}} 
\mathrm{e}^{-2\Omega_{h_1h_2}^{el}(s,b)}=
(2\pi)^2\delta^2\left( \vec{q}\right) + 2\pi \bar{T}_{h_1h_2}(s,q^2),
\nonumber\\
&& \bar{T}_{h_1h_2}(s,q^2)= \int_0^\infty b\,db\, J_0\left( b\sqrt{-q^2}\right)
\left[ 
\mathrm{e}^{-2\Omega_{h_1h_2}^{el}(s,b)}-1
\right]
\nonumber
\end{eqnarray}

\begin{eqnarray}
T_{h_1h_2}^{el}(s,b)&=&\frac{\mathrm{e}^{-2\Omega_{h_1h_2}^{el}(s,b)}-1}{2\mathrm{i}},\nonumber\\
\Omega_{h_1h_2}^{el}(s,b) &=& -\mathrm{i}\,\delta_{h_1h_2}^{el}(s,b),\nonumber\\
\delta^{el}_{h_1h_2}(s,b)&=&\frac{1}{16\pi s}\int_0^\infty d(-t) J_0(b\sqrt{-t}) \delta^{el}_{h_1h_2}(s,t)
\nonumber
\end{eqnarray}

\begin{equation}
\hat{F}_{h}=\mathrm{e}^{(\hat{t}-m_{h}^2)/\Lambda_{h}^2},
\nonumber
\end{equation}

\begin{equation}
M_0^{C,\pi}\left( \{ p \}\right)=
T^{el}_{\pi^+p}(s_{13},t_1)
{\cal P}_{\pi}(\hat{s},\hat{t})
\left[ 
\hat{F}_{\pi}\left( \hat{t}\right)
\right]^2
T^{el}_{\pi^-p}(s_{24},t_2), \nonumber
\end{equation}

\begin{eqnarray}
&& {\cal P}_{\pi}(\hat{s},\hat{t}) =
\left(
\mathrm{ctg}\frac{\pi\alpha_{\pi}(\hat{t})}{2}-\mathrm{i}
\right)\cdot
\frac{\pi\alpha_{\pi}^{\prime}}{2\phantom{\mbox{\large{I}}^2}\!\!\!\!\mathrm{\Gamma}(1+\alpha_{\pi}(\hat{t}))}
\left(
\frac{\hat{s}}{s_0}
\right)^{\alpha_{\pi}(\hat{t})},
\nonumber\\
&& \alpha_{\pi}(\hat{t})=0.7(\hat{t}-m_{\pi}^2)\nonumber
\end{eqnarray}

\begin{equation}
M_0^{C,p}\left( \{ p \}\right)=
T^{el}_{pp}(s_{13},t_1)
{\cal P}_{p}(\hat{s},\hat{t})
\left[ 
\hat{F}_{p}\left( \hat{t}\right)
\right]^2
T^{el}_{\bar{p}p}(s_{24},t_2), \nonumber
\end{equation}

\begin{eqnarray}
&& {\cal P}_{p}(\hat{t}) =
\left(
\mathrm{ctg}\frac{\pi\alpha_{\pi}(\hat{t})}{2}-\mathrm{i}
\right)\cdot
\frac{\pi\alpha_{\pi}^{\prime}}{2\phantom{\mbox{\large{I}}^2}\!\!\!\!\mathrm{\Gamma}(1+\alpha_{\pi}(\hat{t}))}
\left(
\frac{\hat{s}}{s_0}
\right)^{\alpha_{\pi}(\hat{t})},
\nonumber\\
&& \alpha_{\pi}(\hat{t})=0.7(\hat{t}-m_{\pi}^2)\nonumber
\end{eqnarray}

\begin{eqnarray}
&&V_{h_1h_2}(s,q^2)=\int d^2\vec{b} \,\mathrm{e}^{\mathrm{i}\vec{q}\vec{b}}
\sqrt{
1+2\mathrm{i}T_{h_1h_2}^{el}(s,b)}
=\nonumber\\
&& =\int d^2\vec{b}\, \mathrm{e}^{\mathrm{i}\vec{q}\vec{b}} 
\mathrm{e}^{-\Omega_{h_1h_2}^{el}(s,b)}=
 (2\pi)^2\delta^2\left( \vec{q}\right) + 2\pi \tilde{T}_{h_1h_2}\nonumber\\
&& \tilde{T}_{h_1h_2} = \int_0^\infty b\,db\, J_0\left( b\sqrt{-q^2}\right)
\left[ 
\mathrm{e}^{-\Omega_{h_1h_2}^{el}(s,b)}-1
\right]
\nonumber
\end{eqnarray}


\end{document}